\documentclass{aa}
\usepackage[dvips]{graphicx}
\usepackage{txfonts}
\def\kms{\mbox{km~s$^{-1}$}}

\def\Vlsr{$V_{\rm LSR}$}

\def\jykms{~Jy~beam$^{-1}$\,km~s$^{-1}$}

\def\mjy{~mJy~beam$^{-1}$}

\def\jdo{\mbox{$J$=12$\rightarrow$11}}

\def\jdu{\mbox{$J$=2$\rightarrow$1}}
\def\juz{\mbox{$J$=1$\rightarrow$0}}

\def\iras{IRAS 22272+6358A}

\begin{document}

\title{Who Is Eating the Outflow?: High-Angular Resolution Study of an Intermediate-Mass Protostar in L1206}


\author{M.\ T.\ Beltr\'an \inst{1} \and J.\ M.\ Girart \inst{2} \and R.\ Estalella \inst{1}} 

\offprints{M. T. Beltr\'an, \email{mbeltran@am.ub.es}}

\institute{
Departament d'Astronomia i Meteorologia, Universitat de Barcelona, Av. Diagonal 647, 08028 Barcelona, Catalunya, Spain  
\and
Institut de Ci\`encies de l'Espai (CSIC-IEEC), Campus UAB, Facultat de Ci\`encies, Torre C-5, 08193, Bellaterra,
Catalunya, Spain}

\date{Received date; accepted date}

\titlerunning{An Intermediate-Mass Protostar in L1206}
\authorrunning{Beltr\'an et al.}

\abstract
{Up to now only a few intermediate-mass molecular outflows have been studied with enough high-angular
resolution.}
{The aim of this work is to study in detail the intermediate-mass YSO \iras, which is
embedded in L1206, and its molecular outflow, in order to investigate the interaction of the outflow with the dense
protostellar material, and to compare their properties with those of lower mas counterparts.}
{We carried out OVRO observations of the 2.7~mm continuum emission, CO~(\juz), C$^{18}$O(\juz), and HC$_3$N~(\jdo)
in order to map with high-angular resolution the core of L1206, and to derive the properties
of the dust emission, the molecular outflow and the dense protostellar envelope.}
{The 2.7~mm continuum emission has been resolved into four sources, labeled OVRO~1, 2, 3, and 4. The
intermediate-mass Class~0/I object OVRO~2, with a mass traced by the dust emission of 14.2~$M_{\odot}$, is the source associated
with \iras. The CO~(\juz) observations have revealed a very collimated outflow driven by OVRO~2, at a
PA $\simeq 140\degr$, that has a very weak southeastern red lobe and a much stronger northwestern blue
lobe. Photodissociation toward the red lobe produced by the ionization front coming from the bright-rimmed diffuse H{\sc
ii} region could be responsible of the morphology of the outflow.  The spatial correlation between
the outflow and the elongated dense protostellar material traced by HC$_3$N~(\jdo) suggests an
interaction between the molecular outflow and the protostellar envelope. Shocks produced by the
molecular outflow, and possibly by the shock front preceding the ionization front could account for the southern enhancement of HC$_3$N. The properties of the intermediate-mass
protostar OVRO~2 and the molecular outflow are consistent with those of lower mass counterparts. The C$^{18}$O abundance relative to molecular hydrogen estimated
toward OVRO~2 is $3\times10^{-8}$, a value $\sim$6 to 13 times lower than typical abundances
estimated toward molecular clouds. The most plausible explanation for such a difference is CO
depletion toward OVRO~2.}
{}

\maketitle

\section{Introduction}

Molecular outflows are an ubiquitous phenomenon during the earliest stages of star formation for Young
Stellar Objects (YSOs) of all masses and luminosities. During the last decades, many studies have been
devoted to the study and description of the physical properties of  molecular outflows driven by low-mass
protostars and of their embedded driving sources (e.g.\ Bachiller~\cite{bachiller96}; Richer et
al.~\cite{richer00}). In recent years,  high-mass molecular outflows have also been studied in detail, and
many surveys have been carried out toward massive star-forming regions to achieve a more accurate picture of
their morphology and properties (e.g.\ Shepherd \& Churchwell~\cite{shepherd96a}, \cite{shepherd96b}; Zhang
et al~\cite{zhang01}; Shepherd~\cite{shepherd05}). However, although the interest for the lower and, in
particular, for the higher end of the mass spectrum has been growing in recent years, this has not been the
case for intermediate-mass protostars, which have masses in the range $2~M_{\odot}\leq
M_\star\leq10~M_{\odot}$ and also power energetic molecular outflows. In fact, only few  deeply embedded
intermediate-mass protostars are known to date, and only for a very limited number of them their outflows
have been studied at high-angular resolution (e.g.\ NGC 7129: Fuente et al.~\cite{fuente01}; HH288: Gueth et
al.~\cite{gueth01}; IRAS~21391+5802: Beltr\'an et al.~\cite{beltran02}).

\begin{table*}
\caption[] {Parameters of the OVRO observations}
\label{tobs}
\begin{tabular}{lcccrccc}
\hline
&&&\multicolumn{2}{c}{Synthesized beam} \\
\cline{4-5}
\multicolumn{1}{c}{Observation} &
\multicolumn{1}{c}{Configuration} &
\multicolumn{1}{c}{Frequency} &
\multicolumn{1}{c}{{\it HPBW}} &
\multicolumn{1}{c}{PA} &
\multicolumn{1}{c}{Bandwidth} &
\multicolumn{1}{c}{Spectral resolution}& 
\multicolumn{1}{c}{rms noise$^a$}\\
& & \multicolumn{1}{c}{(GHz)} &
\multicolumn{1}{c}{(arcsec)} &
\multicolumn{1}{c}{(deg)} &
\multicolumn{1}{c}{(MHz)} &
\multicolumn{1}{c}{(\kms)}& 
\multicolumn{1}{c}{(mJy~beam$^{-1}$)}
\\
\hline
2.7~mm continuum &L, C &112.53 &$6.3\times4.8$ &$-27$ &4000&$-$ &1.4 \\
CO (\juz) &L, C  &115.27 &$8.0\times6.3$ &$-29$ &22.5 &0.65 &45\\
C$^{18}$O (\juz) &L, C &109.78  &$8.0\times6.6$ &$-32$ &7.5 &0.68 &40\\
HC$_3$N (\jdo) &L, C  &109.17 &$8.1\times6.7$ &$-37$ &7.5 &0.69 &40\\
\hline 

\end{tabular}

(a) For the molecular line observations the 1~$\sigma$ noise is per channel.

\end{table*}

Intermediate-mass protostars are rare in comparison with their low-mass counterparts and tend to be located
at greater distances. The immediate vicinity of intermediate-mass protostars is a very complex environment,
where the extended emission is usually resolved into more than one source when observed at high-angular resolution
(e.g.\ G173.58+2.45: Shepherd \& Watson~\cite{shepherd02}; IRAS~21391+5802: Beltr\'an et
al.~\cite{beltran02}). The molecular outflows driven by intermediate-mass objects are usually more energetic. Thus,
their interaction with the circumstellar gas and dust material surrounding the protostars is expected to be
stronger and more dramatic, disrupting the envelopes and pushing away the dense gas at high velocities.
Although these outflows are, in general, less collimated and more chaotic than those of low-mass stars
(e.g.\ NGC~7129: Fuente et al.~\cite{fuente01}), this could probably be due to observational constraints.
Since these regions are located, on average, further away, these outflows are usually observed with less
linear spatial resolution than nearby low-mass flows. However, when observed with high-angular resolution
their appearance seems to be more collimated and less chaotic (e.g.\ HH~288: Gueth et al.~\cite{gueth01};
IRAS~21391+5802: Beltr\'an et al.~\cite{beltran02}). Therefore, in order to better understand
intermediate-mass stars and compare their morphology and evolution with those of low-mass stars, more
high-angular observations of the dust and gas around  them are needed.

In this paper, we present a detailed interferometric study  of the intermediate-mass protostar \iras\ and of its
molecular outflow. \iras\ is a $1200~L_{\odot}$ (Sugitani et al.~\cite{sugitani89}) YSO located at a distance of
910~pc (Crampton \& Fisher~\cite{crampton74}) and deeply embedded in the bright-rimmed cloud L1206 (BRC~44:
Sugitani et al.~\cite{sugitani91}). This infrared source has no optical counterpart, and at near-infrared
wavelengths, the embedded source has been only seen in scattered light (Ressler \& Shure~\cite{ressler91}). At
far-infrared wavelengths, the source, undetected at 12~$\mu$m, emits most of its energy at 60 and 100~$\mu$m and
has a flux density at 100~$\mu$m greater than that at 60~$\mu$m. This means that its $T(60/100)$ color
temperature is lower than 50~K (38~K; Casoli et al.~\cite{casoli86}), and therefore, that the source is very
embedded, cold, and young. The embedded source has been detected at 2.7 and 2~mm (Wilking et
al.~\cite{wilking89}; Sugitani et al.~\cite{sugitani00}), but not at 2 nor 6~cm wavelengths (Wilking et
al.~\cite{wilking89}; McCutcheon et al.~\cite{mccutcheon91}). Sugitani et al.~(\cite{sugitani89}) have
discovered a CO molecular outflow in the region, although they have only mapped the blueshifted outflow
lobe, since the redshifted lobe was too faint to be mapped with the sensitivity achieved with their  CO (\juz)\,  4-m
radio telescope observations. The core toward \iras\ has also been observed in different high-density tracers,
such as HCO$^+$ (Richards et al.~\cite{richards87}), NH$_3$ (Molinari et al.\cite{molinari96}), CS and
N$_2$H$^+$ (Williams \& Myers~\cite{williams99}), and $^{13}$CO and C$^{18}$O (Ridge et al.~\cite{ridge03}).

\section{Observations}
\label{obs}

Millimeter interferometric observations of L1206 at 2.7~mm were carried out with the Owens Valley Radio
Observatory (OVRO) Millimeter Array of six 10.4~m telescopes in the L (Low) and  C (Compact) configurations, on
2003 May 1 and June 1, respectively. The data taken in both array configurations were combined, resulting in baselines
ranging from 15 to 115~m, which provided sensitivity to spatial structures from about 4$\farcs$8 to 37$''$. The digital
correlator was configured to observe simultaneously the continuum emission and some molecular lines. Details of
the observations are given in Table~\ref{tobs}. The phase center was located at  $\alpha$(J2000) = $22^{\rm h}
28^{\rm h} 51\fs50$,  $\delta$(J2000) = $+64\degr 13' 42\farcs3$. Bandpass calibration was achieved by observing
the quasars 3C84, 3C273, and 3C345. Amplitude and phase were calibrated by observations of the nearby quasar
J1927+739, whose flux density was determined relative to Uranus. The uncertainty in the amplitude calibration is
estimated to be $\sim$$20\%$. The OVRO primary beam is $\sim$$64''$ (FWHM) at 115.27 GHz. The data were calibrated
using the MMA package  developed for OVRO (Scoville et al.~\cite{scoville93}). Reduction and analysis of the data
were carried out using standard procedures in the MIRIAD and GILDAS software packages.  We subtracted the
continuum from the line emission directly in the ({\it u, v})-domain for C$^{18}$O (\juz) and HC$_3$N (\jdo), and 
in the image domain for CO (\juz).

\begin{table*}
\caption[] {Positions, millimeter flux densities, sizes, and masses of the cores detected in L1206}
\label{table_clumps}
\begin{tabular}{lccccccccc}
\hline
&\multicolumn{2}{c}{Position} \\ 
 \cline{2-3} 
&\multicolumn{1}{c}{$\alpha({\rm J2000})$} &
\multicolumn{1}{c}{$\delta({\rm J2000})$} &
\multicolumn{1}{c}{$I^{\rm peak}_\nu$(2.7mm)$^a$} &
\multicolumn{1}{c}{$S_\nu$(2.7mm)$^b$} &
\multicolumn{1}{c}{$\theta^c$} &
\multicolumn{1}{c}{Mass$^d$} &
\multicolumn{1}{c}{$n({\rm H_2})^e$} &
\multicolumn{1}{c}{$N({\rm H_2})^e$} \\
\multicolumn{1}{c}{Core} &
\multicolumn{1}{c}{h m s}&
\multicolumn{1}{c}{$\degr$ $\arcmin$ $\arcsec$} &
\multicolumn{1}{c}{({mJy beam$^{-1}$})} & 
\multicolumn{1}{c}{(mJy)} &
\multicolumn{1}{c}{(arcsec)} &
\multicolumn{1}{c}{($M_\odot$)} &
\multicolumn{1}{c}{(cm$^{-3}$)} &
\multicolumn{1}{c}{(cm$^{-2}$)} \\
\hline
OVRO 1  &22 28 50.49 &+64 13 50.7 &$\phantom{2} 6.8 \pm 1.4$ &$\phantom{1} 9.8 \pm2.0$  &9.3 &1.6 
&$6.5\times10^5$ &$5.5\times10^{22}$\\
OVRO 2$^f$  &22 28 51 41 &+64 13 41.1 &$\phantom{1} 59.2\pm11.8$ &$\phantom{1} 86.0\pm17.2$ &3.5
&$14.2\phantom{1}$
&$1.1\times10^8$ &$3.4\times10^{24}$  \\ 
OVRO 3  &22 28 53.25 &+64 13 32.1 &$10.9\pm2.2$ &$11.0\pm2.2$  &5.4 &1.8  &$3.7\times10^6$ &$1.8\times10^{23}$  \\
OVRO 4  &22 28 53.98 &+64 13 34.5 &$10.7\pm2.1$ &$13.0\pm2.6$   &5.6 &2.2  &$4.0\times10^6$ &$2.0\times10^{23}$  \\
\hline

\end{tabular}
   
  (a) Peak intensity corrected for primary beam response. The uncertainty in the values of the flux density is $\sim
   20\%$. \\
  (b) Integrated flux density corrected for primary beam response. The uncertainty in the values of the flux density is $\sim
   20\%$. \\
  (c) Deconvolved geometric mean of the major and minor axes of the 50\%  of the peak contour. \\  
  (d) Total (gas+dust) circumstellar mass, obtained assuming a dust temperature of 33~K (see Sect. \ref{mass}). \\
  (e) Average H$_2$ volume and column density estimated assuming spherical symmetry and a mean
  molecular mass per H$_2$ molecule of $2.8m_{\rm H}$. \\
  (f) Associated with \iras.
  \end{table*}

\begin{figure}
\centerline{\includegraphics[angle=0,width=8cm]{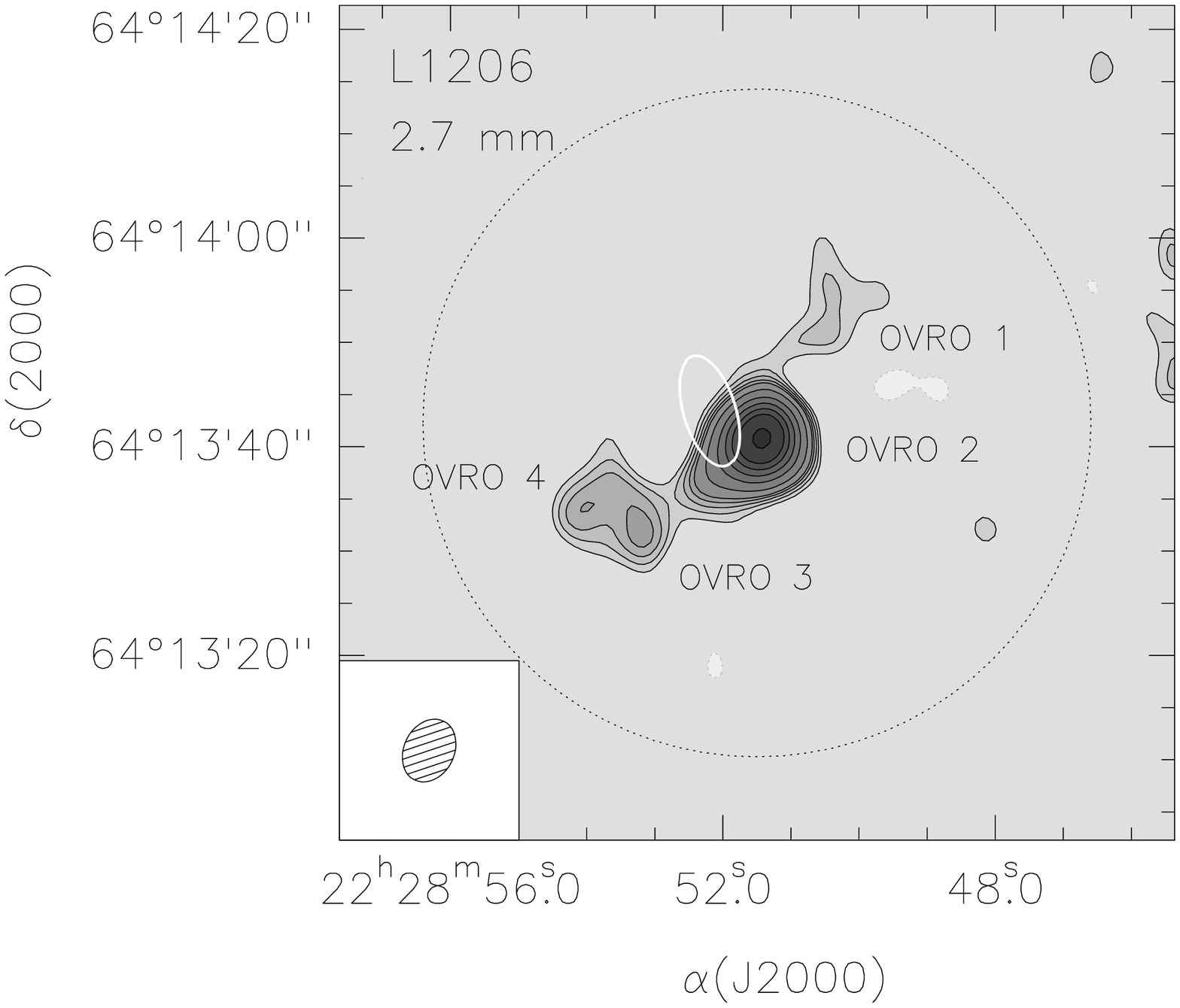}}
\caption{OVRO map of the 2.7~mm continuum emission toward the core of L1206. The synthesized beam is
  $6\farcs3\times4\farcs8$ at $\rm P.A.=-27\degr$, and is drawn in
  the bottom left corner. The rms noise of the map, $\sigma$, is 1.4~mJy\,beam$^{-1}$. 
  The contour levels are $-5~\sigma$, $-3~\sigma$, 3 to 7 times $\sigma$ by 1~$\sigma$, 10~$\sigma$, 15 to 30 times $\sigma$ by
  5~$\sigma$, and 40~$\sigma$. The error ellipse of IRAS~22272+6358A is indicated in white. The dotted circle represents the OVRO primary beam (50\% attenuation level).}
\label{cont}
\end{figure}

\begin{figure*}
\centerline{\includegraphics[angle=-90,width=14cm]{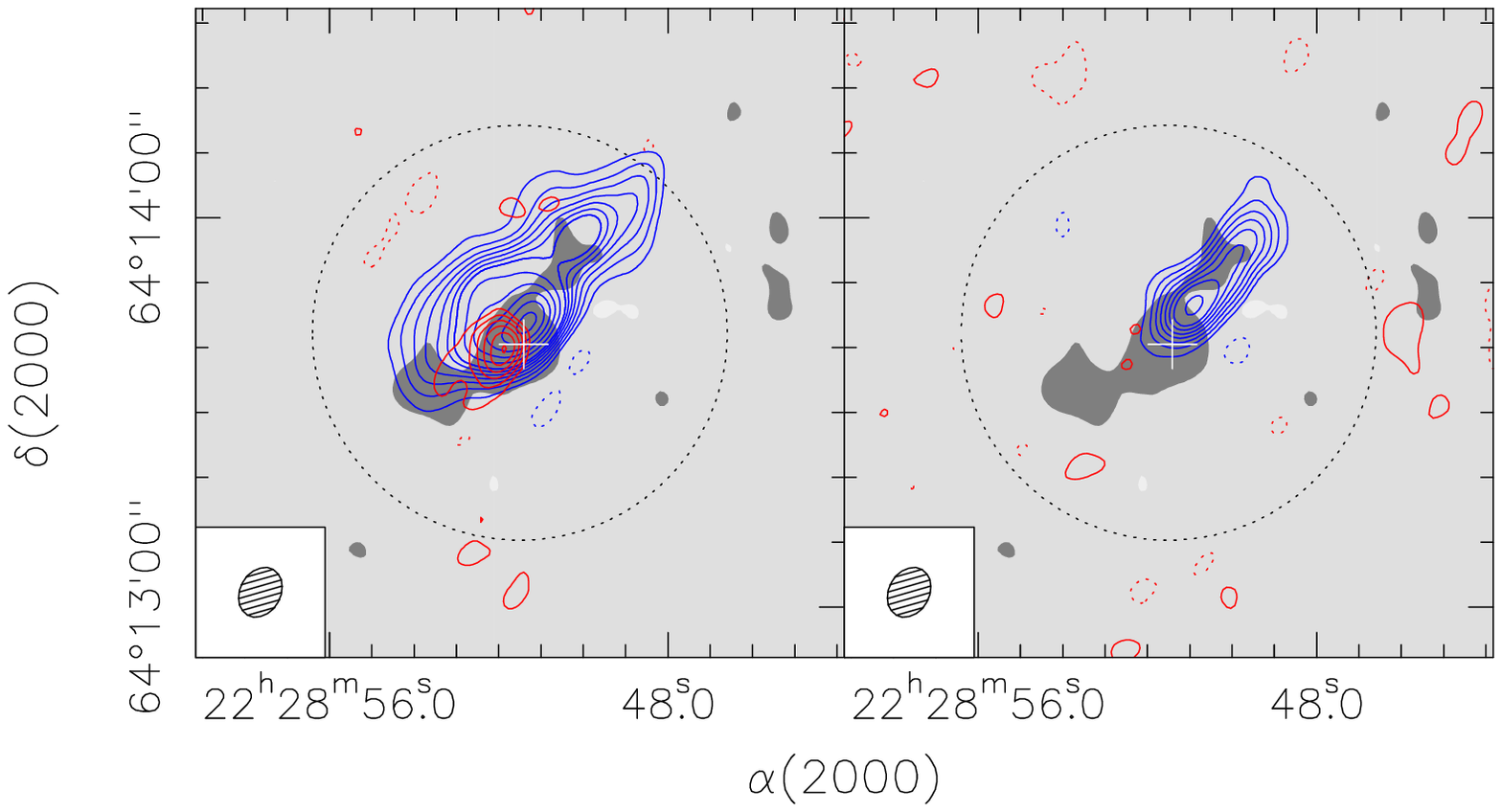}}
\caption{CO (\juz) emission integrated in different velocity intervals, ($-19.5$, $-13.5$) \kms\  for the
low-velocity blueshifted emission {\it(left panel; blue contours)}, ($-8.5$, $-2.5$) \kms\ for the redshifted one
{\it(left panel; red contours)}, ($-30.5$, $-19.5$) \kms\ for the  high-velocity blueshifted emission {\it(right
panel; blue contours)}, and ($-2.5$, +$8.5$) \kms\, for the  high-velocity redshifted one {\it(right panel; red
contours)}, overlaid on the 2.7~mm continuum emission {\it(greyscale)}. Contour levels are $-3~\sigma$,
3 to 15 times $\sigma$ by steps of  2~$\sigma$, and 15 to 35 times $\sigma$ by steps of 4~$\sigma$, where
1~$\sigma$ is 0.5\jykms\ {\it(left panel; blue contours)}, 0.18\jykms\ {\it(left panel; red contours)}, 
0.4\jykms\ {\it(right panel; blue contours)}, and  0.15\jykms\ {\it(right panel; red contours)}. The synthesized
beam is drawn in the bottom left  corner. The dotted circle represents the OVRO primary beam (50\% attenuation
level). The white cross marks the millimeter continuum position of OVRO~2.}
\label{co_outflow}
\end{figure*}

\section{Results}

\subsection{Dust emission}

In Fig.~\ref{cont} we show the OVRO map of the 2.7~mm continuum emission toward the core of L1206, where \iras\ is
embedded. The
continuum dust emission mapped at 2~mm by Sugitani et al.~(\cite{sugitani00}) is resolved out in four
different clumps, labeled OVRO~1, OVRO~2, OVRO~3, and OVRO~4, with the higher  angular resolution of our
interferometric observations. The position, flux density at 2.7~mm,  deconvolved size of the sources,
measured as the geometric mean of the major and minor axis of the 50\%
of the peak contour around each source, the mass of the gas for each source, and the average H$_2$ volume and
column density  are given in
Table~\ref{table_clumps}. The total integrated flux at 2.7mm is $\sim$$120\pm24$ mJy. This value is consistent with
the 2.7~mm expected flux of $\sim 125$~mJy derived from the 2~mm integrated flux of 414~mJy obtained with
the  single dish Nobeyama 45-m telescope (Sugitani et al.~\cite{sugitani00}) for a dust absorption coefficient
proportional to $\nu^2$. Therefore, this indicates that no significant fraction
of the total flux has been filtered out by the interferometer.

The strongest source detected at 2.7~mm in the region is OVRO~2. As can be seen in Fig.~\ref{cont}, this
source is the object associated with the infrared source IRAS 22272+6358A. OVRO~2 is located at the center of
the CO molecular outflow detected in the region (see next section and Fig.~\ref{co_outflow}), which suggests
that it is its driving source. The millimeter emission of OVRO~2 shows two components, a centrally peaked source,
which has a diameter of $\sim$$3\farcs5$ ($\sim$3200 AU) at the 50\% of the dust emission peak, plus an
extended and quite spherical envelope, which has a size of $\sim$$13''$ ($\sim$11800 AU) at the  3~$\sigma$
contour level. By fitting an elliptical Gaussian to the visibility data at the position of the dust
emission peak, we found  a total flux of 78~mJy, and a size of
$3\farcs6\pm0\farcs3\times2\farcs6\pm0\farcs2$ at PA = $-75\degr$. This size corresponds to a linear size of 
$\sim $$3200 \times 2400$~AU. These sizes are consistent with the values found for envelopes around low- and
intermediate-mass protostars (e.g., Hogerheijde et al~\cite{hogerheijde97}, \cite{hogerheijde99}; Looney et
al.~\cite{looney00}; Fuente et al.~\cite{fuente01}; Beltr\'an et al.~\cite{beltran02}). 

The source OVRO~1 is a clump, located $\sim$$11''$ northwest of OVRO~2, elongated along the CO outflow axis
(see Fig.~\ref{co_outflow}). In fact, as can be seen in this figure, OVRO~1 clearly coincides with the the
blueshifted emission lobe, and therefore, it could be dusty material entrained by the outflow. The other two
sources in L1206, OVRO~3 and 4, which are located $\sim$15--18$''$ southeast of OVRO~2, have similar flux
densities ($\sim$11--13 mJy) and deconvolved size ($\sim$5000~AU). Both sources are separated
by only $\sim$$5\farcs3$ ($\sim$5000~AU) and could be sharing a common envelope around them.

\subsection{The molecular outflow: CO emission}

The molecular outflow in L1206 associated with IRAS~22272+6358A, that is, OVRO~2, has been previously studied
through lower angular resolution CO observations by Sugitani et al.~(\cite{sugitani89}). However, since the red
wing of this outflow is so weak, those authors were not able to map the red lobe. In addition, from
their CO blueshifted emission integrated map, it is not clear the direction of the outflow. The observations
reported in this study improve significantly the angular resolution of the outflow maps, revealing the structure
of the molecular outflow in detail, and its direction as well. In addition, we have been able to detect and map
the weak outflow redshifted emission.

\begin{figure}
\centerline{\includegraphics[angle=-90,width=8cm]{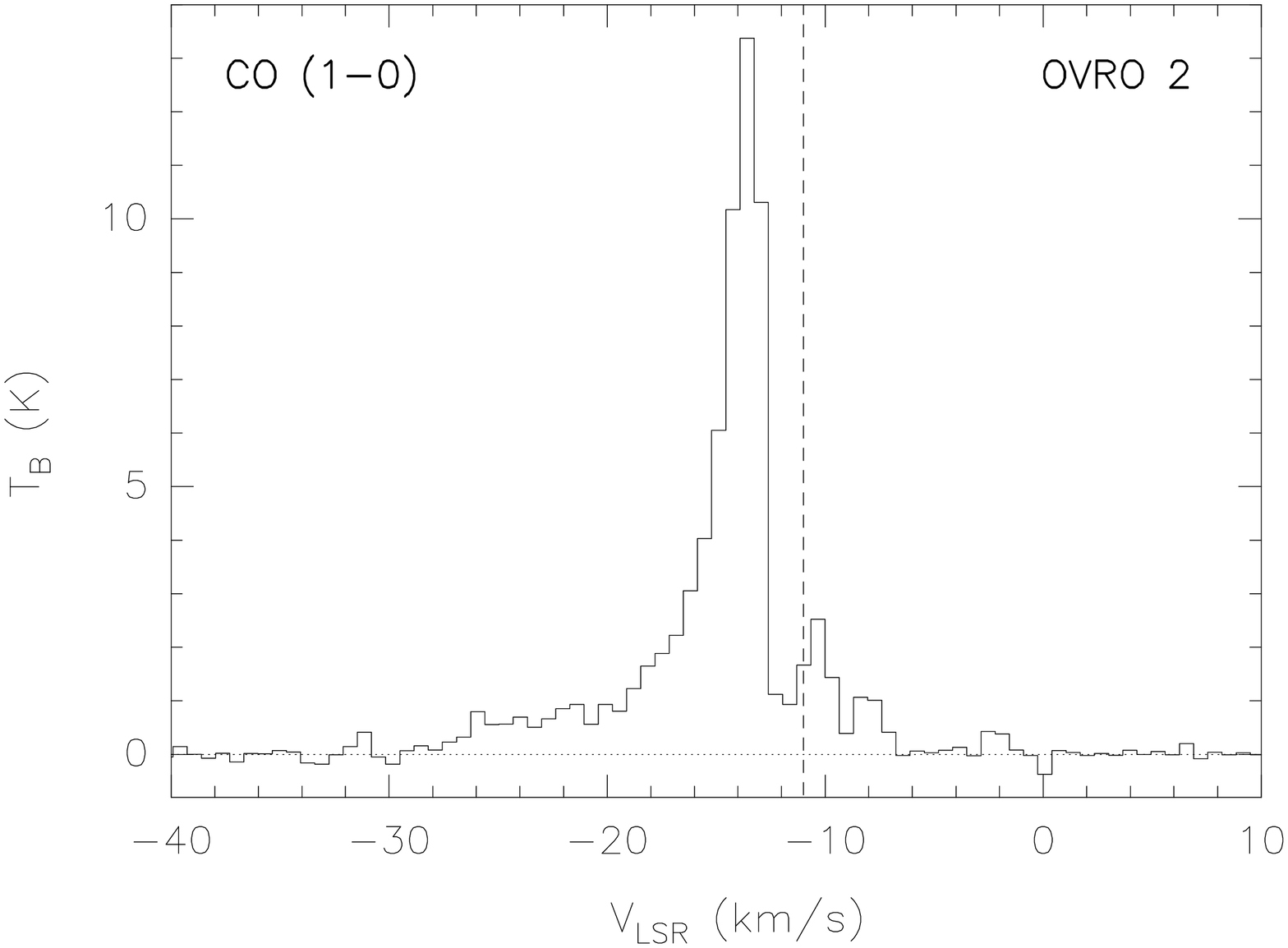}}
\caption{CO (\juz) spectrum obtained at the position of
OVRO~2 in L1206. The continuum has been subtracted. The 1~$\sigma$ level in one channel is 0.10~K. The
conversion factor is 1.82~K/Jy beam$^{-1}$. The dashed vertical line indicates the systemic velocity of
$-11$~\kms.}
\label{co_spectrum}
\end{figure}

\begin{figure}
\centerline{\includegraphics[angle=0,width=7.5cm]{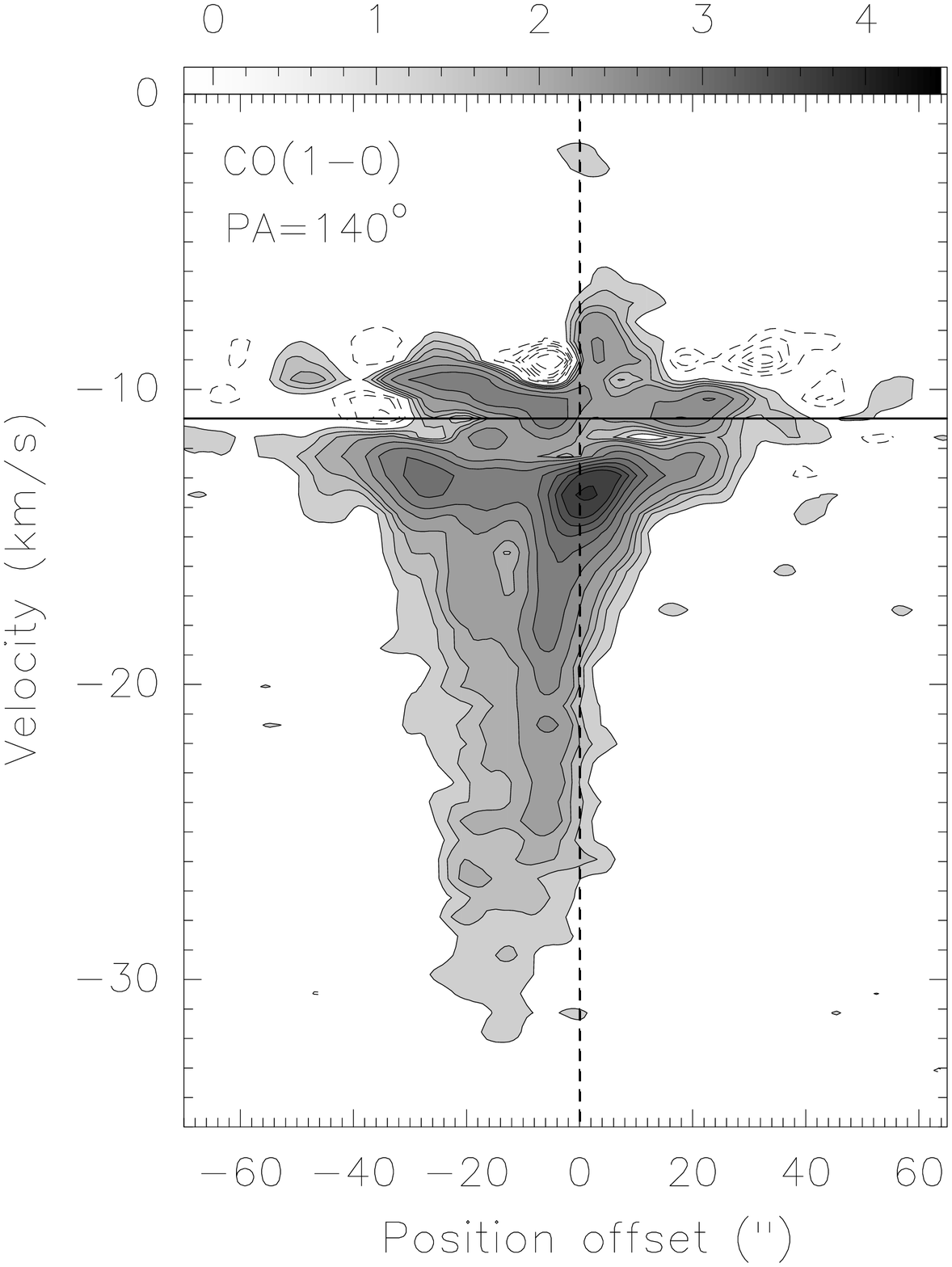}}
\caption{PV plot of the CO (\juz) emission along the major axis, PA$=140\degr$, of the molecular outflow.
The position offset is relative to the projection of the 2.7~mm continuum emission peak onto the outflow axis,
$\alpha$(J2000) = $22^{\rm h} 28^{\rm h} 51\fs52$,  $\delta$(J2000) = $+64\degr 13'  41\farcs7$. Contours
 are $-1.2$, $-0.9$, $-0.72$, $-0.54$, $-0.36$, $-0.18$, 0.18 to 0.9~Jy\,beam$^{-1}$ by steps of
 0.18~Jy\,beam$^{-1}$, 1.2, 1.8, 2.7, 3.6, 4.5, 5.4, and 7.2~Jy\,beam$^{-1}$. The horizontal line marks the
 systemic velocity, $V_{\rm LSR}=-11$~\kms.}
\label{co_cut}
\end{figure}

\begin{figure*}
\centerline{\includegraphics[angle=-90,width=16cm]{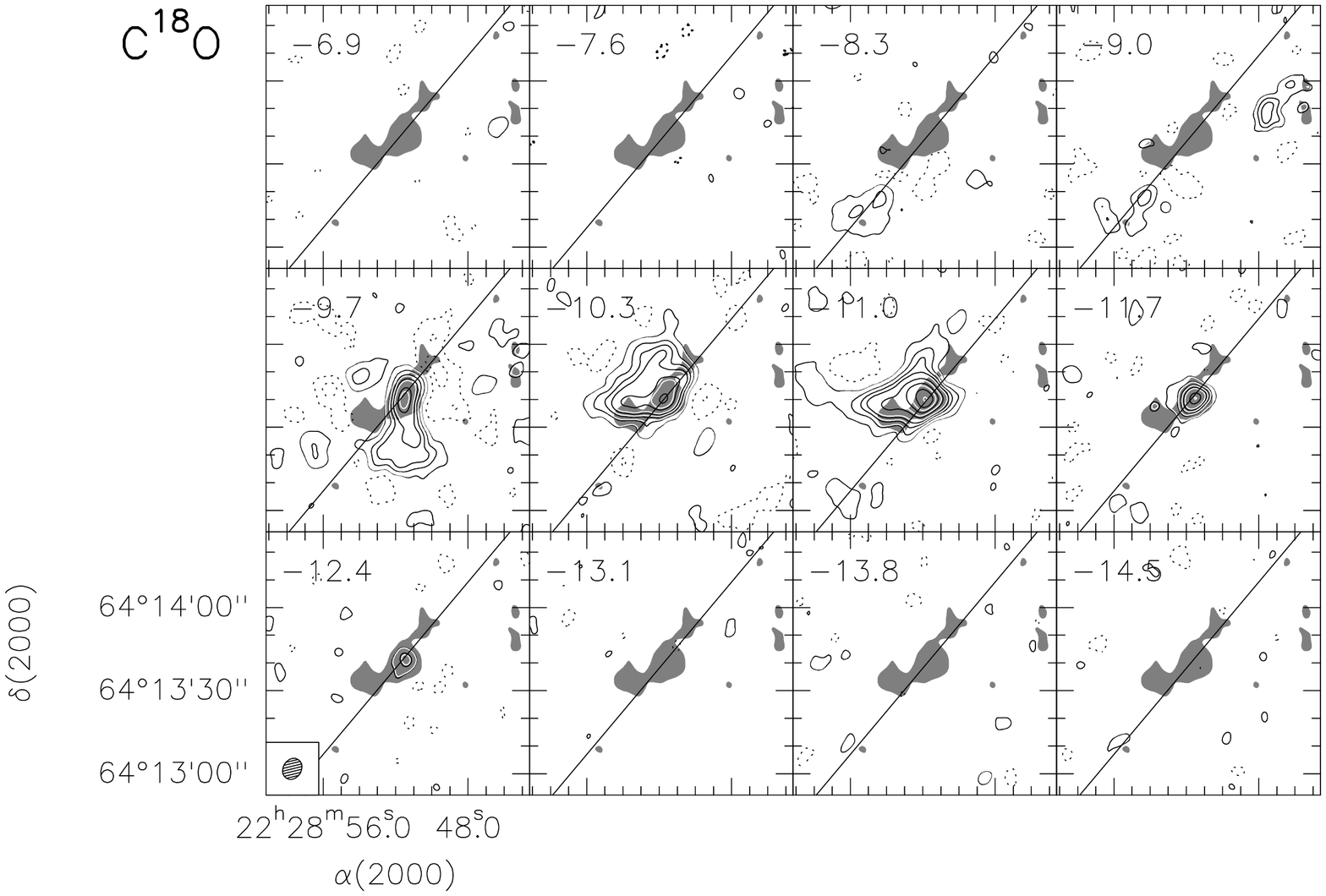}}
\caption{Velocity channel maps of the C$^{18}$O~(\juz) emission {\it (contours)} overlaid on the 2.7~mm continuum
emission {\it (greyscale)}. The $V_{\rm LSR}$ of L1206 is $-11$~\kms. 
The central velocity of each channel is indicated at the upper left corner of each panel. 
The 1~$\sigma$ noise in 1 channel is 40~mJy\,beam$^{-1}$.  Contour levels are $-7$~$\sigma$, $-3$~$\sigma$,  
3 to 15 times $\sigma$ by steps of 4~$\sigma$, and 15 to 39 times $\sigma$ by steps of 8~$\sigma$. The
conversion factor from Jy\,beam$^{-1}$ to K is 1.93. The synthesized beam is drawn in the bottom left 
corner of the bottom left panel. The line outlines the direction of the CO outflow (see
Fig.~\ref{co_outflow}).} 
\label{c18o_channel}
\end{figure*}

Figure~\ref{co_outflow} shows the maps of the integrated CO (\juz) emission in two different blueshifted and
redshifted velocity intervals. The systemic velocity, $V_{\rm LSR}$, reported in the literature is roughly
$-10$~\kms\, (Richards~\cite{richards87}; Sugitani et al.~\cite{sugitani89}; Wilking et al.~\cite{wilking89};
Wouterloot \& Brand~\cite{wouterloot89}; Williams \& Myers~\cite{williams99}). However, from our fits to the
C$^{18}$O (\juz) and  HC$_3$N (\jdo) spectra, we have estimated a systemic velocity $V_{\rm LSR} \simeq
-11$~\kms\ (see next sections), which is the value used in this paper. Regarding the line widths in the region,
which have been estimated from CO and CS and isotopomers, HCO$^+$, and N$_2$H$^+$ lines, the values derived are
in the range 0.7 to 3.0~\kms\, (Richards~\cite{richards87}; Wilking et al.~\cite{wilking89}; Wouterloot \&
Brand~\cite{wouterloot89}; Molinari et al.~\cite{molinari96}; Williams \& Myers~\cite{williams99}; Ridge et
al.~\cite{ridge03}). Therefore, taking this into account, we conservatively considered the CO ambient cloud velocity
interval to be ($-13.5$, $-8.5$) \kms, that is, $-11\pm2.5$ \kms. And for the outflow,  we took the ($-19.5$,
$-13.5$) \kms\ velocity interval for the low-velocity blueshifted emission, ($-8.5$, $-2.5$) \kms\ for the
redshifted one, ($-30.5$, $-19.5$) \kms\ for the  high-velocity blueshifted emission, and ($-2.5$, +$8.5$)
\kms\, for the redshifted one. As can be seen in  Fig.~\ref{co_outflow} the molecular outflow is
elongated in a direction with PA = $140\degr$ and has a very weak redshifted emission. In fact, the southeastern
red lobe is only visible at low outflow velocities. The outflow is clearly centered at the position of OVRO~2,
which seems to be its driving source. Figure~\ref{co_spectrum} shows the CO (\juz) spectrum toward the position
of OVRO~2, where a blueshifted wing that extends to very high velocities, and a weak  redshifted wing are
clearly visible . In this figure a strong self-absorption observed for the central velocity channels is also
visible. This self-absorption feature is probably due to the filtering out of part of the extended emission by
the interferometer, although we cannot exclude opacity effects as well, and it makes impossible to study the
cloud structure with CO (\juz). The direction of the outflow and the presence of a red wing are also confirmed
by the position-velocity (PV) cut done along the major axis of the outflow ($140\degr$), and centered in the
projection of the OVRO~2 dust emission peak position onto the outflow axis  (Fig.~\ref{co_cut}).

This picture of the molecular outflow supports the scenario drawn by Ressler \& Shure~(\cite{ressler91}) from
near-infrared polarimetry, photometry, and imaging of the core of L1206. According to these authors, the
near-infrared northwestern object named A-1, which is composed of scattered light, is the blueshifted lobe of
the jet pointed slightly toward us, while the weaker southeastern object A-2, also composed of scattered light,
is the redshifted lobe pointed away. The position of these two lobes coincides with the CO blueshifted and
redshifted lobes, respectively. The protostar that is powering the outflow has to be heavily embedded, as is not detected at
near-infrared wavelengths. Ressler \& Shure~(\cite{ressler91}) propose that  the driving source is a Class~I
object with an optically thick disk lying almost edge-on to the line-of-sight, and located at $\alpha$(J2000) =
$22^{\rm h} 28^{\rm h} 51\fs54$,  $\delta$(J2000) = $+64\degr 13'  42\farcs3$, which is roughly coincident with
the 2.7~mm OVRO~2 position (see Table~\ref{table_clumps}).

\subsection{C$^{18}$O emission}

\begin{figure}
\centerline{\includegraphics[angle=0,width=7.5cm]{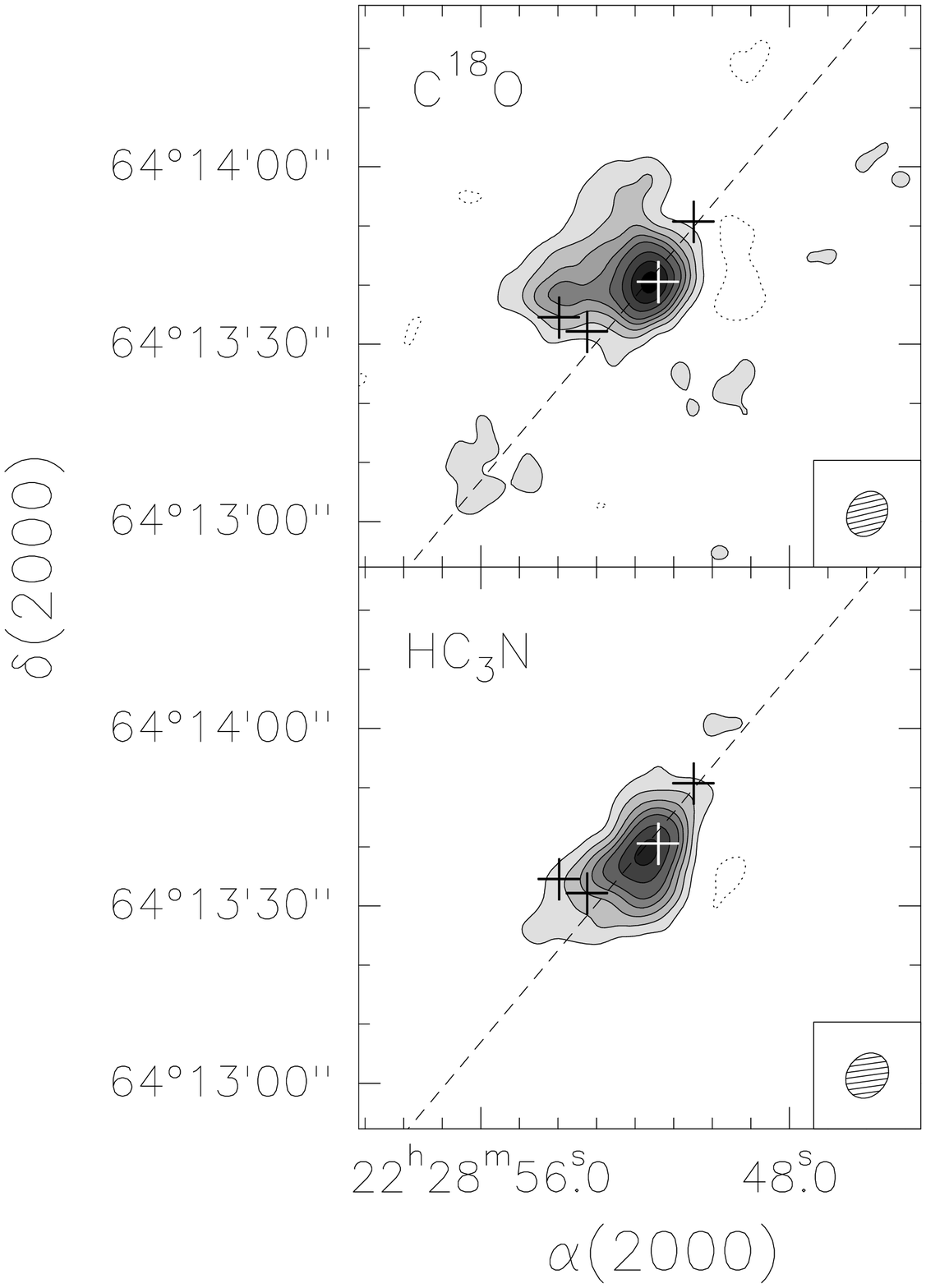}}
\caption{{\it (Top)} Integrated intensity map of the C$^{18}$O (\juz)  emission over the velocity interval
($-12.4$,$-8.3$)~\kms\ toward L1206. Contour levels are $-0.36$, 0.36 to 1.8\jykms\ by steps of 0.36\jykms,
and 1.8 to 3.6\jykms\ by steps of 0.6\jykms. {\it (Bottom)} Integrated intensity map of the  HC$_3$N~(\jdo)
emission over the velocity interval ($-11.7$,$-9.7$)~\kms. Contour levels are $-0.3$, 0.3 to 1.5\jykms\ by
steps of 0.3\jykms, and 1.5 to 2.5\jykms\ by steps of 0.5\jykms. The crosses show the positions of the 2.7~mm
continuum sources detected in the region. The dashed line indicates the direction of the CO outflow. The synthesized beam is
drawn in the bottom right corner of each panel.}
\label{average}
\end{figure}

Figure~\ref{c18o_channel} shows the velocity channel maps for the  C$^{18}$O~(\juz) emission around the systemic
velocity, $V_{\rm LSR}\simeq -11$~\kms, overlaid on the continuum emission. The emission integrated over the
central channels, velocity interval ($-12.4$,$-8.3$)~\kms, is shown in the top panel of Fig.~\ref{average}.  The
emission is dominated by a central compact component that peaks at the position of the continuum
intermediate-mass continuum source OVRO~2, as it can be clearly seen for the channel maps at velocities of
$-11.7$ and $-12.4$~\kms\ (see Fig.~\ref{c18o_channel}). Therefore, such emission is tracing the innermost part
of the protostellar envelope. For velocities between $-9.7$ and $-11.0$~\kms\ the C$^{18}$O emission shows an
additional extended component elongated north and east of OVRO~2 and also in part associated with the continuum
sources OVRO~3 and 4. As can be seen in the channel and integrated maps, there is also some redshifted emission,
visible at velocities of $-8.3$ and $-9.0$~\kms, which is located in the same direction as the CO redshifted
lobe (see  Fig.~\ref{co_outflow}) but much more southeastwards than the emission traced by CO. The total
integrated intensity of the central emission is $\sim$15\jykms, and  has a deconvolved size, measured as the
geometric mean of the major and minor axes of the 50\% of the peak  contour of the gas, of $11\farcs4$ or $\sim
10400$~AU.

\subsection{HC$_3$N emission}
\label{hc3n}

\begin{figure*}
\centerline{\includegraphics[angle=-90,width=16cm]{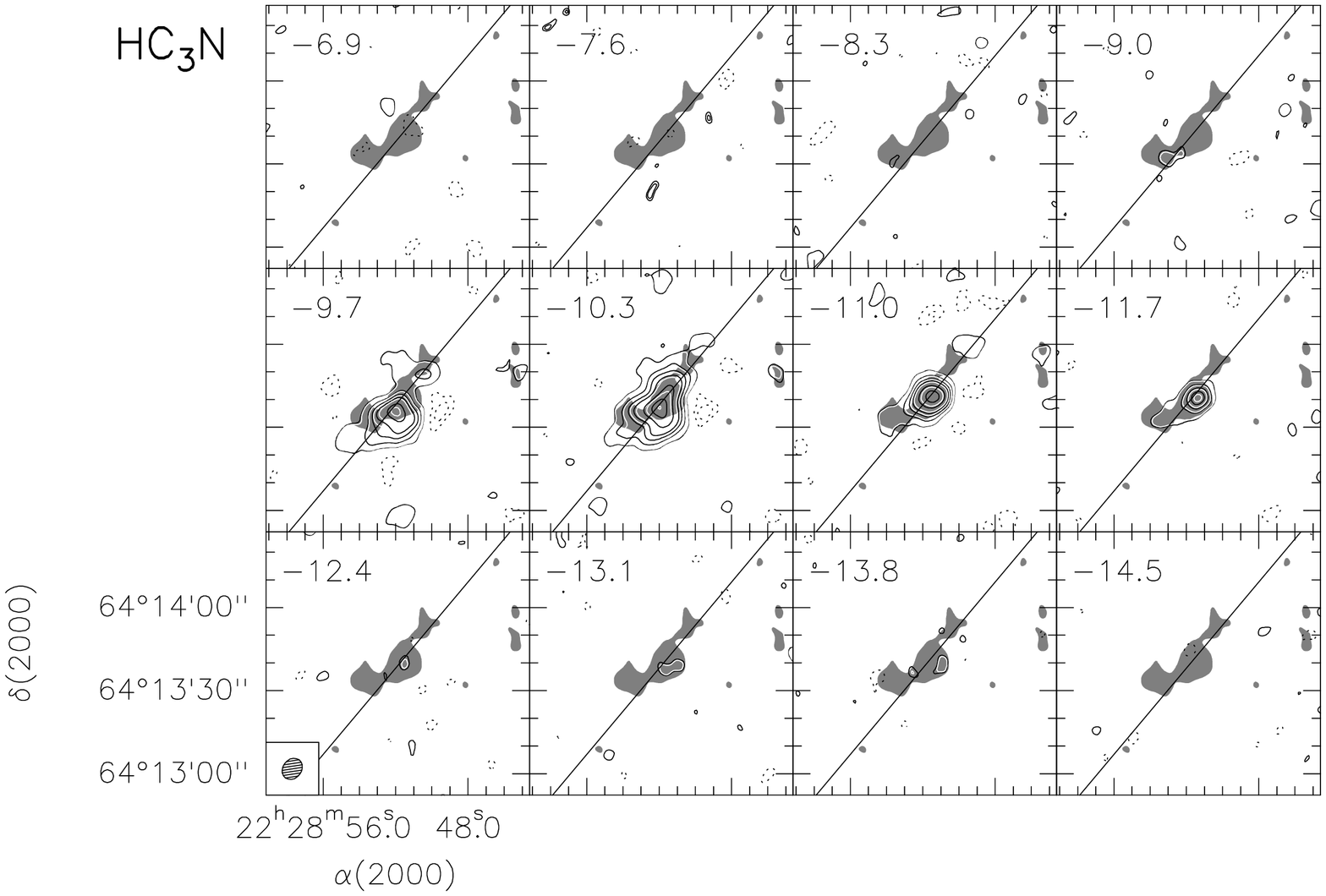}}
\caption{Velocity channel maps of the HC$_3$N~(\jdo) emission {\it (contours)} overlaid on the 2.7~mm continuum
emission {\it (greyscale)}. The $V_{\rm LSR}$ of L1206 is $-11$~\kms.  
The central velocity of each channel is indicated at the upper left corner of each panel. 
The 1~$\sigma$ noise in 1 channel is 40~mJy\,beam$^{-1}$.  Contour levels are $-5$~$\sigma$,$-3$~$\sigma$,  
3  to 15 times $\sigma$ by steps of 4~$\sigma$, and 15 to 31 times $\sigma$ by steps of 8~$\sigma$. The
conversion factor from Jy\,beam$^{-1}$ to K is 1.88. The synthesized beam is drawn in the bottom left 
corner of the bottom left panel. The line outlines the direction of the CO outflow (see
Fig.~\ref{co_outflow}).}
\label{hc3n_channel}
\end{figure*}

The HC$_3$N (\jdo) transition has a high critical density ($8\times10^5$~cm$^{-3}$: Chung et
al.~\cite{chung91}), which makes it an even better tracer of high density regions than e.g.\ CS~(\juz) and
CS~(\jdu). Figure~\ref{hc3n_channel} shows the velocity channel maps for the HC$_3$N~(\jdo) emission toward the
core of L1206, around the systemic velocity, $V_{\rm LSR}\simeq -11$~\kms, overlaid on the continuum emission.
The emission integrated over the velocity interval ($-11.7$,$-9.7$)~\kms\ is shown in the bottom panel of
Fig.~\ref{average}. The gas emission at velocities of $-11.0$ and $-11.7$~\kms\ is very compact and peaks at the
position of the continuum intermediate-mass protostar OVRO~2. At these velocities an emission tail is also visible
toward the southeast, which is associated with the continuum sources OVRO~3 and 4. For the redshifted
velocities $-9.7$ and $-10.3$~\kms, the emission shows a less compact and more elongated morphology (see
Figure~\ref{hc3n_channel}). The emission peak of the gas at these two redshifted velocities is located
southeastwards of OVRO~2, in the same direction as the  the CO redshifted lobe (see  Fig.~\ref{co_outflow}). As
can be seen in the channel and integrated emission maps, the gas emission is elongated along the direction of
the CO molecular outflow.  This morphological correlation  between the outflow and dense protostellar material
is similar to the case of the low-mass Class~0 source in L1157 (Beltr\'an et al.~\cite{beltran04a}), mapped in
the same molecular line HC$_3$N~(\jdo). However, for  L1206, our OVRO observations did not provide us with
enough angular resolution to trace in HC$_3$N the apparent base of the outflow conical cavities as traced by the
CO, and to more clearly map the ``cross-like'' pattern characteristic of the interaction between the molecular
outflow and the protostellar envelope. The integrated HC$_3$N emission has a flux density $\sim$10.5\jykms, and
a deconvolved size of $8\farcs8$ or $\sim 8000$~AU.

\begin{figure}
\centerline{\includegraphics[angle=-90,width=8.5cm]{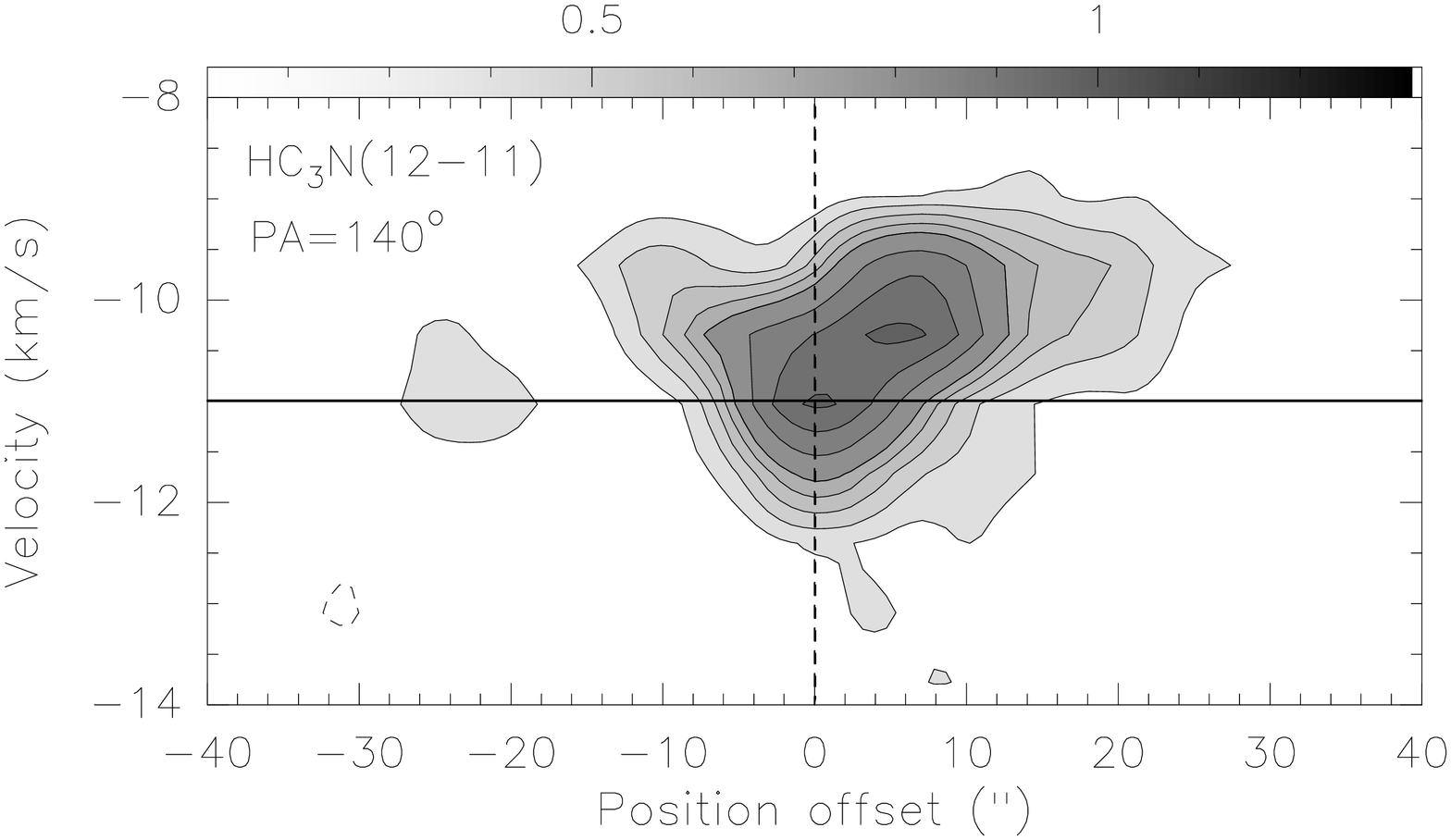}}
\caption{PV plot of the HC$_3$N (\jdo) emission along the major axis, PA$=140\degr$, of the molecular
outflow. The position offset is relative to the projection of the 2.7~mm continuum emission peak onto the
outflow axis, $\alpha$(J2000) = $22^{\rm h} 28^{\rm h} 51\fs52$,  $\delta$(J2000) = $+64\degr 13' 
41\farcs7$. Contours are $-0.12$, 0.12 to 0.6~Jy\,beam$^{-1}$ by steps of 0.12~Jy\,beam$^{-1}$, and 0.6 to
1.32~Jy\,beam$^{-1}$ by steps of 0.24~Jy\,beam$^{-1}$. The horizontal line marks the systemic velocity,
$V_{\rm LSR}=-11$~\kms.}
\label{hc3n_cut}
\end{figure}

The kinematics of HC$_3$N toward the core of L1206 can be seen in the PV cut done along the direction of
OVRO~2 molecular outflow (see Fig.~\ref{hc3n_cut}). The emission shows two peaks, one at the position of the
continuum source, which corresponds to offset position zero, at the systemic velocity, and the other one $\sim
5\farcs5$ southeast from the central position, at a velocity of $\sim -10.3$~\kms.  In addition to this
redshifted emission peak, an excess of emission, or emission tail, is also clearly visible extending further away than the CO
redshifted lobe (up to $\sim$$27''$ or $\sim$24600~AU southeastwards of OVRO~2), and reaching
velocities up to $\sim$$-9$~\kms.

\section{Analysis and Discussion}

\subsection{Dust emission}
\label{mass}

The masses of the cores given in Table~\ref{table_clumps} have been estimated assuming that the dust emission is
optically thin, by using 
\begin{equation}
M_{\rm clump}=\frac{g\,S_\nu\,d^2}{\kappa_\nu\,B_\nu(T_{\rm d})},
\end{equation}
where ${S_\nu}$ is the flux density, $d$ is the distance to the source, ${\kappa_\nu}$ is the dust mass opacity coefficient,
$g$ is the gas-to-dust ratio, and $B_\nu(T_{\rm d})$ is the Planck function for a blackbody of dust temperature $T_{\rm d}$,
all measured at $\nu=112.5$~GHz. We adopted $\kappa_{112.5}$ = 0.20~cm$^2$\,g$^{-1}$ ($\kappa_{0}$ = 1~cm$^2$\,g$^{-1}$ at
250~GHz: Ossenkopf \& Henning~\cite{ossenkopf94}),  $g=100$, and $T_{\rm d}= 33$~K for all the sources. Estimates of  $T_{\rm
d}$ have been obtained by fitting a grey-body to the IRAS fluxes at 60 and 100~$\mu$m and to the sum of fluxes of the 4 clumps
at 2.7~mm  We have neglected the IRAS fluxes at 12 and 25~$\mu$m in the fit. The reason for this choice is that two components
are known to be present in the SEDs of luminous YSOs (see Sridharan et al.~\cite{srid02}): one associated with compact, hot gas
and dominating the 12 and 25~$\mu$m fluxes; the other due to colder material, contributing to the 60 and 100~$\mu$m emission.
The latter is the one of interest to us because we want to estimate the temperature of the low-temperature dust in the
circumstellar envelope of the sources. Adopting a dust absorption coefficient proportional to $\nu^2$, we found $T_{\rm
d}\simeq 33$~K. This is the same value obtained by Sugitani et al.~(\cite{sugitani00}) fitting the fluxes at 2~mm, 100 and
60~$\mu$m, with a dust absorption coefficient proportional to $\nu^{1.8}$. 

The mass estimated for OVRO~2 is 14.2~$M_{\odot}$, while for the other sources the mass is 1.6, 1.8, and
2.2~$M_{\odot}$ for OVRO~1, OVRO~3, and OVRO~4, respectively. Therefore, OVRO~2, which has a circumstellar
mass consistent with the YSO being an intermediate-mass source,  accounts for most of the mass toward the
core of L1206. The mass estimates of the other sources suggest that they are low-mass YSOs. We also
estimated the average H$_2$ volume density, $n({\rm H_2})$, and column density, $N({\rm H_2})$, of the sources
by assuming spherical symmetry and a mean molecular mass per H$_2$ molecule of $2.8m_{\rm H}$. The values are given in
Table~\ref{table_clumps}.

\subsection{Physical parameters of the CO outflow}
\label{co_prop}

\begin{table*}
\caption[] {Properties of the CO outflow}
\label{tco}
\begin{tabular}{lccccc}
\hline
&\multicolumn{1}{c}{$v^{a}$}&
\multicolumn{1}{c}{$M^{b}$}&
\multicolumn{1}{c}{$P^{c}$}&
\multicolumn{1}{c}{$E^{c}$}&
\multicolumn{1}{c}{$\dot P^{c}$}
\\
\multicolumn{1}{c}{Lobe}&
\multicolumn{1}{c}{(\kms)}&
\multicolumn{1}{c}{($10^{-2} M_\odot$)}&
\multicolumn{1}{c}{($M_\odot$ \kms)}&
\multicolumn{1}{c}{($10^{43}$ erg)}&
\multicolumn{1}{c}{($M_\odot \, \mbox{km s}^{-1}$\,yr$^{-1}$)}
\\
\hline
Blue &[$-30.5, -13.5$] &8.3  &0.60  &4.78   &$7.5\times10^{-5}$\\
Red$^{d}$ &[$-8.5, -2.5$]    &0.2  &0.01 &0.03  &$2.3\times10^{-6}$ \\
Total &                &8.5  &0.61 &4.81  &$7.7\times10^{-5}$ \\
\hline

\end{tabular}

(a)  Range of outflow velocities. Momenta and kinetic energies are calculated relative to
the cloud velocity, which is taken to be $V_{\rm LSR}=-11$~\kms. \\
(b) Assuming an excitation temperature of 24~K. \\
(c) Velocity not corrected for inclination. \\
(d) The outflow has only been detected for this velocity range.  
\end{table*}

Assuming that the CO emission is in LTE and is optically thin, the mass of the gas associated with the 
outflow detected in the
region can be calculated by using 
\begin{eqnarray} 
\label{co_mass}
\nonumber
\bigg(\frac{M}{M_{\sun}}\bigg)=2.3\times10^{-5}\,\frac{T_{\rm ex} + 0.93}{{\rm
exp}(-5.59/T_{\rm ex})} \, \bigg(\frac{d}{\rm kpc}\bigg)^2 \\
\times \int{\bigg(\frac{S_\nu}{\rm Jy}\bigg)\,\bigg(\frac{dv}{\rm km\,s^{-1}}\bigg)},
\end{eqnarray} 
where $T_{\rm ex}$ is the CO excitation temperature, $d$ is the distance in kpc, and $S_\nu$ is the line flux
density measured in Jy. An excitation temperature of 24~K has been estimated from the CO (\jdu) antenna
temperature, $T_{\rm A}^ \ast=15.4$~K, obtained by Wilking et al.~(\cite{wilking89}), which corresponds to
$T_R=18.6$~K (Kutner \& Ulich~\cite{kutner81}) for an efficiency $\eta_c=1$ and a forward scattering and spillover
efficiency $\eta_{\rm fss}=0.83$. We assumed an [H$_2$]/[CO] abundance ratio of 10$^4$ (e.g.\ Scoville et
al.~\cite{scoville86}), and a mean atomic mass
per H atom $\mu=1.4$. In Table~\ref{tco} we give the masses, the CO momentum, the kinetic energy, and the
momentum rate in the outflow. We also report the range of outflow velocities for the outflow. Note that due to
the fact that some emission is resolved out by the interferometer, and the absorption by the ambient cloud,  the
masses calculated should be considered as lower limits. There is also the possibility that the CO is optically
thick in portions of the flow,  although the correction for opacity should be small, as one would expect that
for high outflow velocities the material is optically thin. Finally, it should be taken into account the
integration range chosen, which could make that part of the outflow at low velocities is not counted. Note that
the momentum, kinetic energy, and momentum rate in the outflow have not been corrected for the (unknown)
inclination angle, $i$, of the flow with respect to the plane of the sky. In case of correcting for inclination,
the velocities should be divided by $\sin i$. The values of the mass, momentum, kinetic energy, and momentum
rate are smaller than the values estimated for other intermediate-mass molecular outflows, such as
IRAS~20050+2720 (Bachiller et al.~\cite{bachiller95}), IRAS~21391+5802 (Beltr\'an et al.~\cite{beltran02}),
HH288 (Gueth et al.~\cite{gueth01}), L1641S (see references in Anglada~\cite{anglada95} and Wu et
al.~\cite{wu04}), or NGC~2071 (Snell et al.~\cite{snell84}), as can be seen in Table~\ref{toutflows}, and are more
consistent with the values derived for low-mass outflows (e.g.\ Cabrit \& Bertout~\cite{cabrit92};
Anglada~\cite{anglada95}; Bontemps et al.~\cite{bontemps96}; Lee et al.~\cite{lee00}, \cite{lee02}). The reason
for this could be the weakness of the redshifted lobe of the OVRO~2 flow, which is hardly contributing to the
mass, momentum and energy of the molecular outflow. A possible explanation for the weakness of the redshifted
lobe is given in Sect.~\ref{rim}.

\begin{table*}
\caption[] {Properties of intermediate-mass outflows}
\label{toutflows}
\begin{tabular}{lcccc}
\hline
&\multicolumn{1}{c}{$M^{b}$}&
\multicolumn{1}{c}{$P^{c}$}&
\multicolumn{1}{c}{$E^{c}$}&
\multicolumn{1}{c}{$\dot P^{c}$}
\\
\multicolumn{1}{c}{Outflow}&
\multicolumn{1}{c}{($M_\odot$)}&
\multicolumn{1}{c}{($M_\odot$ \kms)}&
\multicolumn{1}{c}{($10^{45}$ erg)}&
\multicolumn{1}{c}{($M_\odot \, \mbox{km s}^{-1}$\,yr$^{-1}$)}
\\
\hline
 IRAS 20050+2720$^{a}$ &1.7 &25.4\phantom{1} &7.7 &$\sim 5.0\times10^{-3}\phantom{13}$\\
 IRAS 21391+5802$^{b}$ &\phantom{1}0.14 &3.6 &1.2 &$1.4\times10^{-3}$ \\
 HH288$^{c}$ &11 &385 &6.7 &$1.4\times10^{-2}$ \\
 L1641S$^{d}$ &1.4 &37 &0.3 &$5.3\times10^{-4}$ \\
 NGC 2071$^{e}$ &10 &81 &8.5 &$5.1\times10^{-3}$ \\
\hline

\end{tabular}

(a) The parameters correspond to the whole emission of the different outflow lobes detected in 
IRAS~20050+2720 (Bachiller et al.~\cite{bachiller95}). \\
(b) Beltr\'an et al.~(\cite{beltran02}). \\
(c) Because of the difficulty to treat each outflow separately the parameters correspond to the whole
emission, thus including the two outflows detected in HH228 (Gueth et al.~\cite{gueth01}). \\
(d) Anglada~(\cite{anglada95}) and Wu et al.~(\cite{wu04}).\\
(e) Snell et al.~(\cite{snell84}).
\end{table*}

\subsection{Envelope mass from gas emission}
\label{gas_emission}

One of the major problems when using the gas emission to estimate the mass of the molecular core is the
uncertainty of the gas abundance relative to molecular hydrogen. Therefore, due to these uncertainties, instead of measuring the mass of the gas toward the intermediate-mass protostar OVRO~2 from
the C$^{18}$O or HC$_3$N emission, it is better to estimate the fractional abundance of  C$^{18}$O and
HC$_3$N.  In order to do that, we integrated the gas emission in the same  area ($\sim$162~arcsec$^2$) used
to estimate the mass of the gas from the continuum dust emission, and assumed that the mass toward OVRO~2
derived from the  HC$_3$N is the same as that derived from the 2.7~mm continuum emission; that is,
14.2~$M_{\odot}$. Following the derivation of Scoville et al.~(\cite{scoville86}), and assuming optically thin
emission, the C$^{18}$O beam averaged column density is given by 
\begin{equation}
\label{c18o_density} 
\langle N \rangle =4.6\,\,10^{13}\, \frac{(T_{\rm ex}+0.878)}{e^{-5.27/T_{\rm
ex}}}\int{T_B\,dv} ~~{\rm cm}^{-2},
\end{equation}
where $T_{\rm ex}$ is the excitation temperature, and $\int{T_B dv}$ is the integrated brightness temperature
of the C$^{18}$O~(\juz) emission in K\,km\,s$^{-1}$. Assuming an excitation temperature $T_{\rm ex}=24$~K,
the estimated C$^{18}$O~(\juz) abundance relative to molecular hydrogen is $3\times10^{-8}$ toward OVRO~2, a 
value that is $\sim$6 to 13 times lower than typical fractional abundances estimated toward molecular
clouds, 1.7--4$\times10^{-7}$ (Frerking et al.~\cite{frerking82}; Kulesa et al.~\cite{kulesa05}). Even if one
takes into account the uncertainties up to a factor of 5 introduced in the mass estimates by the different
dust opacity laws used, the value of the abundance derived toward OVRO~2 would be lower than the  typical
fractional abundances. In particular, it would be clearly inconsistent with the abundance of 4$\times10^{-7}$
derived by Kulesa et al.~(\cite{kulesa05}) in the $\rho$ Ophiuchi molecular cloud. The excitation temperature
value adopted cannot account for the abundance difference, as the abundance estimate is not significantly
affected by the excitation temperature (the abundance would be 2.7$\times10^{-8}$ for $T_{\rm ex}=20$~K, and
4$\times10^{-8}$ for $T_{\rm ex}=40$~K). One should also take into account opacity effects, in case the 
C$^{18}$O emission is optically thick. In such a case the gas mass would be higher and the C$^{18}$O~(\juz)
abundance relative to molecular hydrogen higher. In case the opacity of the gas is $\tau\gtrsim 3$, the main
beam brightness temperature can be given by $T_{\rm mb} \simeq f[J(T_{\rm ex})-J(T_{\rm bg})]$, where $f$ is
the beam filling factor and $T_{\rm bg}$ is the background temperature. For OVRO~2,  $T_{\rm mb}$ is 2.94~K
(see Table~\ref{table_lines}), and  $T_{\rm ex}=24$~K. Therefore, the filling factor would be  $f=0.14$,
suggesting that the emission is clumpy. However, if the C$^{18}$O emission was clumpy, such clumps should be
visible in the maps done with uniform weighting or with negative robust parameter, and this is not the case.
This means that opacity can be ruled out as the responsible of the mass difference. Therefore, the most 
plausible explanation to  account for the low C$^{18}$O~(\juz) abundance appears to
be CO depletion toward OVRO~2. 

Regarding the HC$_3$N emission, the uncertainties in the HC$_3$N abundance relative to molecular hydrogen
(see e.g.\ Hasegawa et al.~\cite{hasegawa86}) can make the mass estimates to vary up to 3 orders of
magnitude. Chung et al.~(\cite{chung91}) estimate abundances of (2--7) $\times10^{-10}$ in massive dense
cores, while that of a low-mass dense core, L1551, has been estimated to be 2$\times10^{-9}$. Similar
abundances have been estimated by J{\o}rgensen et al.~(\cite{jorgensen04}) for a sample of Class~0 and
Class~I low-mass protostellar envelopes. The average values determined are $3.5\times10^{-10}$ and
$3.1\times10^{-9}$, respectively. However, abundances as low as $\sim$$3\times10^{-11}$ have been estimated 
for a sample of 19 low and massive molecular clouds (Vanden Bout et al.~\cite{vandenbout83}), and as high as
$\sim$$10^{-8}$ for hot cores associated with intermediate- and high-mass stars (de Vicente et
al.~\cite{devicente00}; Mart\'{\i}n-Pintado et al.~\cite{martin05}). Therefore, as done before with the C$^{18}$O
emission, we estimated the fractional abundance of HC$_3$N toward OVRO~2. Following the derivation of Scoville et al.~(\cite{scoville86}), and assuming optically thin
emission, the HC$_3$N beam averaged column density is given by
\begin{equation}
\label{hc3n_density} 
\langle N \rangle =4.2\,\,10^{10}\, \frac{(T_{\rm ex}+0.073)}{e^{-34.06/T_{\rm
ex}}}\int{T_B\,dv} ~~{\rm cm}^{-2},
\end{equation}
where $T_{\rm ex}$ is the excitation temperature, and $\int{T_B dv}$ is the integrated brightness temperature
of the HC$_3$N~(\jdo) emission in K\,km\,s$^{-1}$. Assuming an excitation temperature $T_{\rm ex}=24$~K, the
estimated  HC$_3$N abundance relative to molecular hydrogen is $7\times10^{-11}$ toward OVRO~2, a value in
the lower end of the range of fractional abundances estimated toward molecular clouds.


\begin{table*}
\caption[] {C$^{18}$O (\juz), and HC$_3$N (\jdo) line and physical parameters toward the
continuum peak position of OVRO~2 in L1206}
\label{table_lines}
\begin{tabular}{lcccccc}
\hline
\multicolumn{1}{c}{Molecule \&} &
\multicolumn{1}{c}{\Vlsr} &
\multicolumn{1}{c}{FWHM}  &
\multicolumn{1}{c}{$T_{\rm B}$} &
\multicolumn{1}{c}{$\int{T_{\rm B}\,{\rm d}V}$} &
\multicolumn{1}{c}{$\theta^a$} &
\multicolumn{1}{c}{$M_{\rm vir}^b$}
\\
\multicolumn{1}{c}{Transition} &
\multicolumn{1}{c}{(km s$^{-1}$)} &
\multicolumn{1}{c}{(km s$^{-1}$)} &
\multicolumn{1}{c}{(K)} &
\multicolumn{1}{c}{(K km s$^{-1}$)} &
\multicolumn{1}{c}{(arcsec)} &
\multicolumn{1}{c}{($M_\odot$)} 
\\
\hline
C$^{18}$O (\juz) &$-10.74\pm0.02$ &$2.34\pm0.04$ &$2.94\pm0.31$ &$7.31\pm0.13$ &$8.8$ &10.6--14.2 \\
HC$_3$N (\jdo) &$-10.86\pm0.03$ &$1.83\pm0.07$ &$2.68\pm0.10$ &$5.22\pm0.17$ &$11.4$ &13.4--17.8 \\
\hline
\end{tabular}

(a)  Deconvolved geometric mean of the major and minor axes of the 50\% of
 the peak contour of the gas. \\
(b) Estimated from Eq.~\ref{virial} for typical density distributions with $p=2.0$--1.5.
\end{table*}

\begin{figure}
\centerline{\includegraphics[angle=0,width=8cm]{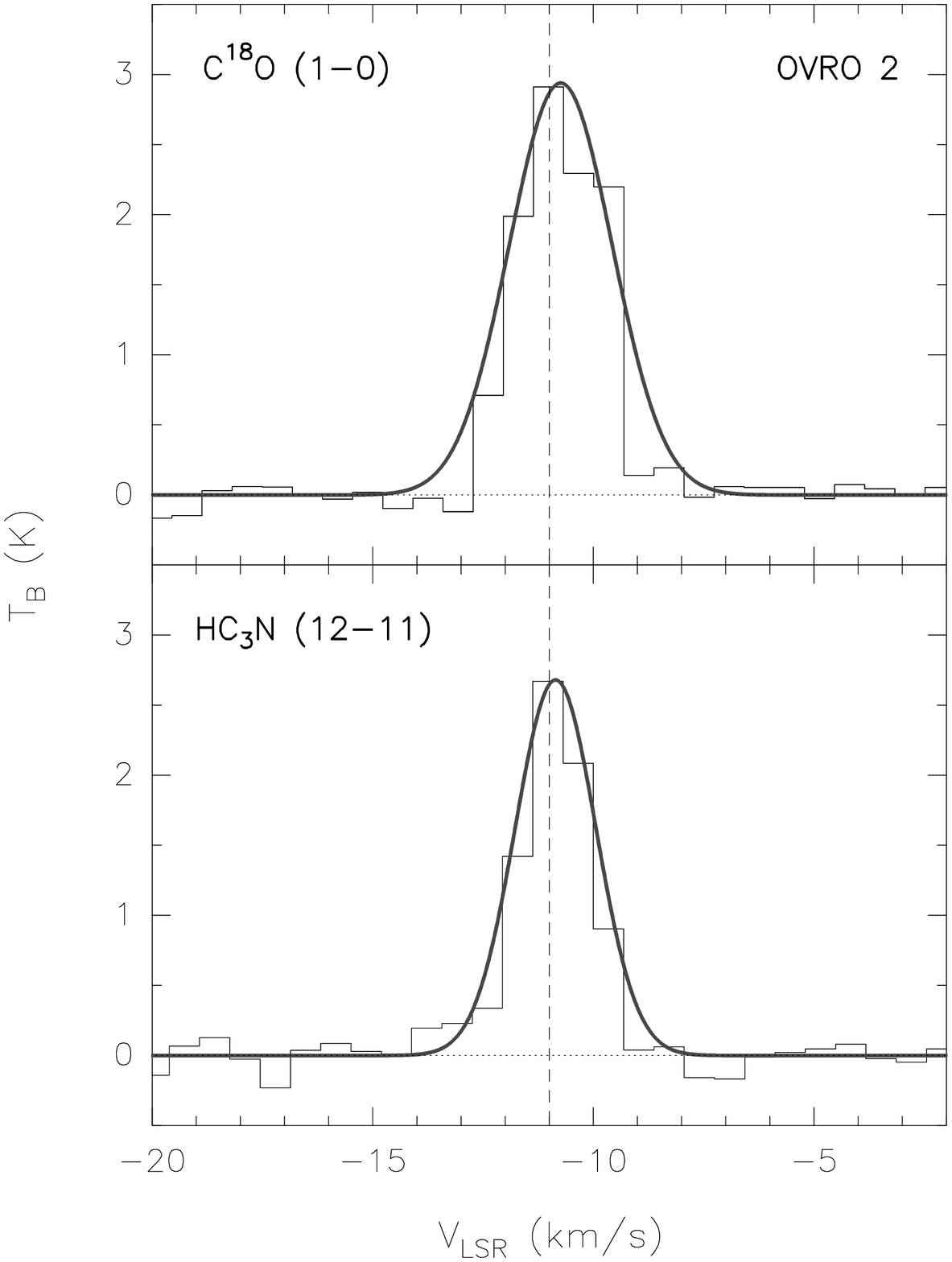}}
\caption{C$^{18}$O (\juz) ({\it top}) and HC$_3$N (\jdo) ({\it bottom}) spectra obtained at the position of
OVRO~2 in L1206. The continuum has been subtracted. The 1~$\sigma$ level in one channel is 0.10~K. The
conversion factor is 1.93~K/Jy beam$^{-1}$ for
C$^{18}$O and 1.88~K/Jy beam$^{-1}$ for HC$_3$N. The thick grey profiles are the Gaussian fits to the
spectra.  The dashed vertical line indicates the systemic velocity of
$-11$~\kms.}
\label{lines}
\end{figure}

\subsection{Physical parameters of the dense core toward OVRO~2}
\label{gas_prop}

Table~\ref{table_lines} lists the fitted parameters for C$^{18}$O (\juz) and HC$_3$N (\jdo) toward the
2.7~mm continuum peak position of OVRO~2. The spectra and the corresponding Gaussian fits are shown in
Fig.~\ref{lines}. In Table~\ref{table_lines} we also give the deconvolved size, and the virial mass
estimated assuming a spherical clump with a power-law density distribution $\rho\propto r^{-p}$, and
neglecting contributions from magnetic field and surface pressure. In such a case the virial mass can be
computed from the expression (see e.g.\ MacLaren et al.~\cite{maclaren88}): 
\begin{equation}  
\bigg(\frac{{M_{\rm vir}}}{M_\odot} \bigg) = 0.305 \,\, \frac{5-2p}{3-p} \,
\bigg(\frac{d}{{\rm kpc}}\bigg) \bigg(\frac{\theta}{{\rm arcsec}}\bigg) 
\bigg(\frac{\Delta V}{{\rm km~s}^{-1}}\bigg)^2,  
\label{virial}  
\end{equation} 
where $p$ is the density distribution power-law index, $d$ is the distance, $\theta$ is the deconvolved size of
the source, and ${\Delta V}$ is the line width. The line widths, which have been estimated from the Gaussian
fits to the spectra, coincide with the line widths estimated from the second order moment maps.  The virial
masses estimated from C$^{18}$O and HC$_3$N  are 13.4--17.8~$M_{\odot}$, and 
10.6--14.2~$M_{\odot}$, respectively, for typical density distributions with $p=2.0$--1.5. Such values are
consistent with the mass derived from the continuum. However, the mass of the central protostar has not been taken into account in this calculation.
Therefore, in fact, the total mass (circumstellar plus protostar) of OVRO~2 is probably higher than the virial
mass, suggesting that this object is likely undergoing collapse, as one would expect from a very young stellar
object.

\subsection{Intermediate-mass protostars and their outflows}

The millimeter continuum emission around \iras\ has been resolved into four different clumps (see
Fig.~\ref{cont}): an intermediate-mass source, OVRO~2, and three less massive and smaller objects, OVRO~1,
OVRO~3, and OVRO~4. This situation resembles very closely the scenario around IRAS~21391+5802, which  is located
in the bright-rimmed cloud IC1396N. In this case, the millimeter and centimeter emission have been resolved into an
intermediate-mass source, BIMA~2, which is the YSO associated with IRAS~21391+5802, surrounded by two
(Beltr\'an et al.~\cite{beltran02}) or maybe three (Beltr\'an et al.~\cite{beltran04b}) much less massive YSOs,
with different morphologies and properties.  As already mentioned, OVRO~2 is the source associated with \iras,
has the strongest millimeter emission, and is also the most massive object toward the core of  L1206. The
source is associated with OH and CH$_3$OH maser emission (MacLeod et al.~\cite{macleod98}) but not with H$_2$O
maser emission (Palla et al.~\cite{palla91}). The source is neither associated with continuum emission at 2 and
6~cm (Wilking et al.~\cite{wilking89}; McCutcheon et al.~\cite{mccutcheon91}). OVRO~2, as seen in previous
sections, is driving a powerful molecular outflow, which has an estimated kinematical age of the order of a few
10$^4$ years, which suggests that the powering source is a very young stellar object. The source, which has not been detected at
12~$\mu$m by IRAS, has a cold dust temperature, $T_{\rm d} \simeq 33$~K, and its millimeter emission, which
accounts for most of the dust emission in the region,  exhibits an extended component that is consistent with an
envelope surrounding the source. The circumstellar mass, $M\simeq 14.2~M_{\odot}$, is consistent with the masses
in the range 3.5--30~$M_{\odot}$ found around other intermediate-mass protostars (Fuente et al.~\cite{fuente01};
Gueth et al.~\cite{gueth01}; Shepherd \& Watson~\cite{shepherd02}), and it is considerably higher than the
envelope masses of 1.4--2.3~$M_{\odot}$ found by Bontemps et al.~(\cite{bontemps96}) for low-mass Class~0
sources. Therefore, these properties indicate that OVRO~2 is an extremely young, deeply embedded
intermediate-mass protostar, with morphology and properties that do not differ significantly from the properties
of low-mass counterpart protostars. In fact, the source, which has not been detected at $J, H, K, L$, or $L'$
bands (Ressler \& Shure~\cite{ressler91}), has been classified as a heavily extinguished Class~I object with an
optically thick almost edge-on circumstellar disk. Pezzuto et al.~(\cite{pezzuto02}) also classify \iras\
as a Class~I object  based on its [60--100]~$\mu$m and [100--170]~$\mu$m colour temperatures, although the low spatial
resolution of the Long Wavelength Spectrometer (LWS) on board of the Infrared Space Observatory can cause a high
uncertainty in the measured fluxes due to background emission or source confusion inside the LWS beam. On the
other hand, taken into account the criterion for Class~0 given by Andr\'e et al.~(\cite{andre93}), $L_{\rm
bol}/L_{\rm submm} \lesssim 200$, Sugitani et al.~(\cite{sugitani00}) classify \iras\ as a Class 0-like
object since its ratio $L_{\rm bol}/L_{\rm submm}$  is 140. Therefore, OVRO~2 seems to be a very young
intermediate-mass object in transition state between Class~0 and I.

In order to further compare the properties of the intermediate-mass YSO OVRO~2 with the low-mass case we also
checked whether this object is consistent with some correlations between source and outflow properties found
for low-mass young objects. As already mentioned in Sect.~\ref{co_prop}, although $L_{\rm bol}$ and the dust
continuum emission toward OVRO~2 indicate that the source is an intermediate-mass object, the properties of
the molecular outflow are more consistent with those derived for low-mass protostars. Therefore, we checked
for the correlation between the circumstellar envelope mass and the momentum rate in the CO outflow given by
Bontemps et al.~(\cite{bontemps96}) for low-mass embedded YSOs.  For this correlation, Bontemps et
al.~(\cite{bontemps96}) correct the observed momentum rate of the CO outflow, $\dot P_{\rm obs}$, by a factor
10, $\dot P\simeq 10\times \dot P_{\rm obs}$, in order to take into account projection and optical depth
effects. After applying the same correction factor to the momentum rate of the OVRO~2 outflow, we found that 
OVRO~2 agrees well with the correlation, and also with the intermediate-mass protostar IRAS~21391+5802
(Beltr\'an et al.~\cite{beltran02}). Bontemps et
al.~(\cite{bontemps96}) also find a correlation between the normalized outflow momentum rate or outflow
efficiency, $\dot P\,c/L_{\rm bol}$, and the normalized envelope mass, $M/L_{\rm bol}^{0.6}$. For \iras, the
estimated bolometric luminosity is  $L_{\rm bol}\simeq1200$~$L_{\odot}$ (Sugitani et al.~\cite{sugitani89}).
However, this estimate includes the whole region around the IRAS catalogue position. Thus, in order to derive a
bolometric luminosity for OVRO~2, we used the relationship between the momentum rate and the bolometric
luminosity given by  Bontemps et al.~(\cite{bontemps96}), and we inferred a luminosity of $L_{\rm bol}\simeq
580~L_{\odot}$. By using this luminosity we found that the source has an outflow efficiency consistent with
those of low-mass Class~I sources, and also with that of IRAS~21391+5802, if one estimates the bolometric
luminosity of the latter from the outflow momentum rate by using the Bontemps et al.~(\cite{bontemps96})
correlation. Both intermediate-mass sources, OVRO~2 and IRAS~21391+5802, fit well the correlation of outflow
efficiency and envelope mass found by Bontemps et al.~(\cite{bontemps96}), and both lie in the limit between
Class~0 and I region of this diagram. Furthermore, another correlation fitted well by both sources is the one 
between the radio continuum luminosity at centimeter wavelengths and the momentum rate of the outflow not
corrected for inclination obtained by Anglada~(\cite{anglada96}). This correlation is in agreement with the
predictions of a simple model of shock ionization in a plane-parallel geometry (Curiel~\cite{curiel87},
\cite{curiel89}), which would be able to produce the required ionization in thermal
radio jets.

Intermediate-mass outflows have been proposed to be more complex and with more chaotic morphologies than those
found toward low-mass protostars (e.g.\ NGC~7129: Fuente et al.~\cite{fuente01}). However, this does not seem to
be the case of the OVRO~2 molecular outflow in L1206. In addition, unlike the  NGC~7129~FIRS1 outflow, the
collimation of the OVRO~2 outflow is very high, even at low velocities, similar to the intermediate-mass outflow
driven by BIMA~2 in IC1396N (Beltr\'an et al.~\cite{beltran02}) or to the low-mass protostellar flow in HH~211
(Gueth \& Guilloteau~\cite{gueth99}). As already discussed by Beltr\'an et al.~(\cite{beltran02}) for the BIMA~2
outflow in IC1396N, given that the outflows are usually more energetic for higher mass objects, that the dust
emission is often resolved into more than one object, and that intermediate- and high-mass sources are embedded
in larger amounts of material, it is to be expected that the interactions between the high-velocity gas and the
circumstellar material will be more dramatic, disrupting and pushing more material. Therefore, the complexity of
the molecular outflows driven by intermediate-mass protostars is likely a result of the more complex
protostellar environment itself. 

\subsection{Interaction of the molecular cloud with the ionized bright-rimmed cloud}
\label{rim}

The  most striking feature of the morphology of the molecular outflow powered by OVRO~2 is the weakness and
small size of the redshifted lobe as compared to the blueshifted one (see Fig.~\ref{co_outflow}). Or,
alternatively, the practically lack of outflowing material south of the YSO OVRO~2. The existence of unipolar,
or primarily one-sided,  molecular outflows have also been reported for other low- and intermediate-mass
star-forming regions (e.g.\ NGC~2024/FIR5: Richer et al.~\cite{richer89}; HH~46-47: Chernin \&
Masson~\cite{chernin91}; L1641-S3: Stanke et al.~\cite{stanke00}; HH~300: Arce \& Goodman~\cite{arce01}). Taking
into account that the molecular outflow consists of ambient gas that is swept up from a position close to where
it is observed,  a possible explanation for the morphology of the OVRO~2 outflow could be that the redshifted
lobe breaks out of the molecular cloud. That is, the redshifted wind passes into a region where there is no, or
little, molecular material to be swept up. A similar scenario has been suggested to explain the near
unipolarity of the HH~46-47 molecular outflow (Chernin \& Masson~\cite{chernin91}).

\begin{figure}
\centerline{\includegraphics[angle=0,width=8.1cm]{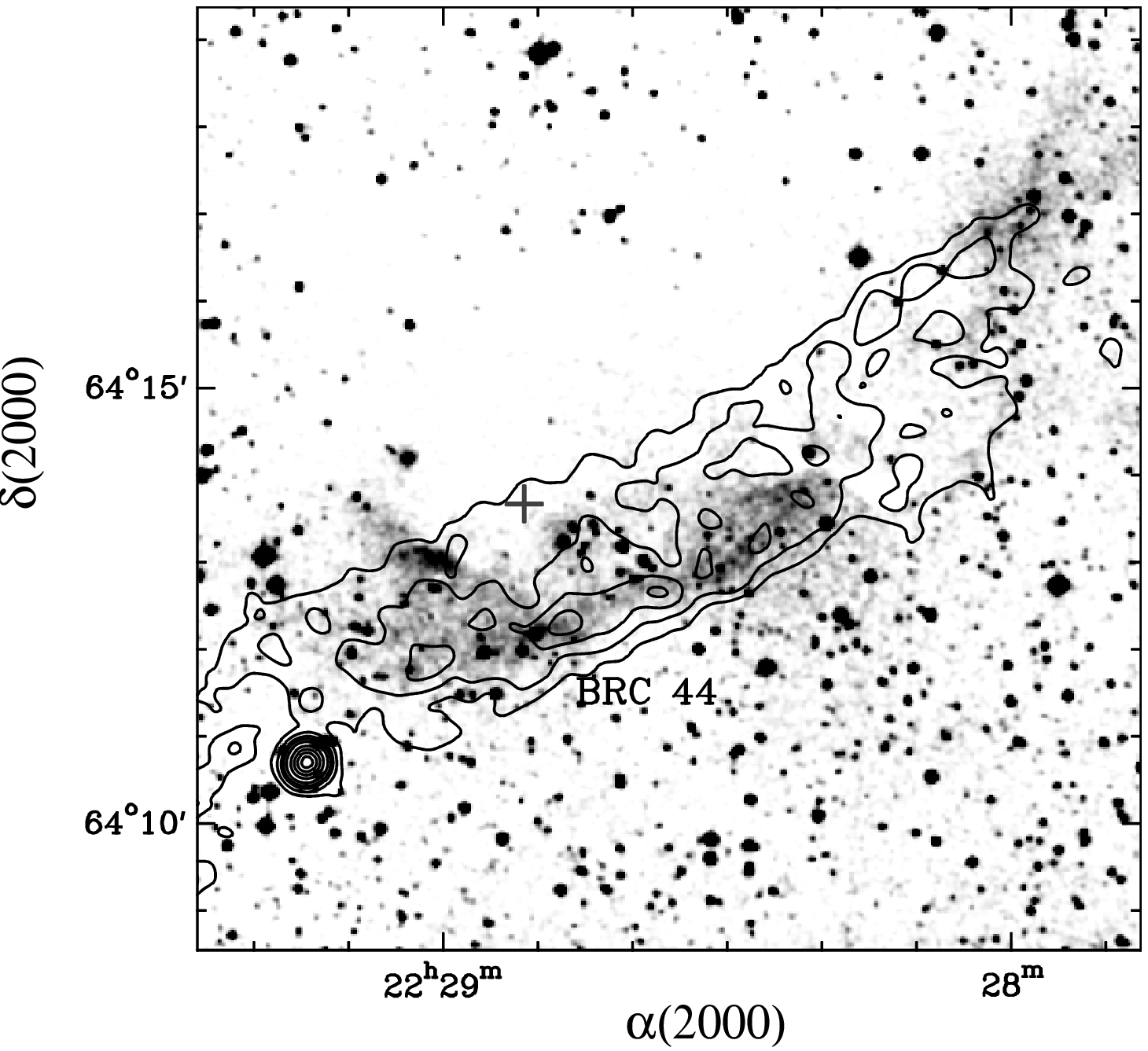}}
\caption{VLA map of the 6~cm emission {\it (contours)} overlaid with an optical image from the Digitized  Sky
Survey 2  {\it (colours)} of BRC~44. The contours are 0.2, 0.4, 0.7, 1.0, 1.5, 2.5, 3.5, 4.5, 5.5, and 6.5~\mjy. 
The cross marks the position of the 2.7~mm continuum source OVRO~2.}
\label{photo}
\end{figure}

Figure~\ref{photo} shows the superposition of the VLA\footnote{The National Radio Astronomy Observatory is a
facility of the National Science Foundation operated under cooperative agreement by Associated Universities,
Inc.} 6~cm emission (archive data) of the bright-rimmed cloud (BRC~44) on the red image of the Digitized Sky
Survey 2 toward L1206. The 6~cm emission is spatially coincident with the L1206 dark cloud. The optical
bright rim, which is a diffuse H{\sc ii} region, coincides with the southern part of the  6~cm emission.
This suggests that the  H{\sc ii} region is located in the background with respect to  L1206. Therefore, a
more plausible explanation for the morphology of the molecular outflow powered by OVRO~2, could be
photodissociation of the redshifted (southern) lobe. The ionization-shock front probably lies very close to
the southern redshifted outflow, and thus, any redshifted (southern) ejected material could be destroyed by
the intense radiation field at the ionization front. Photodissociation has also been proposed as an
explanation of the unipolarity of the NGC~2024/FIR5 outflow (Richer et al.~\cite{richer89}). In order to
check the validity of this possible scenario, we have estimated the internal pressure of the molecular gas
in the L1206 cloud and compared it with the Ionized Boundary Layer (IBL) pressure. The IBL pressure divided
per Boltzmann's constant, $P_{\rm i}/k_{\rm B}$, estimated from the NRAO/VLA sky survey (NVSS) data at 20~cm
for this bright-rimmed cloud is $8.5\times10^6$~cm$^{-3}$ K (Morgan et al.~\cite{morgan04})\footnote{After this paper went to press,  we
realized that there is an error in the Morgan et al.~(\cite{morgan04}) paper (L.\ K.\ Morgan 2006, private
communication), and that the value quoted for the IBL pressure of the cloud should be
lower.  This brings the pressure difference in the cloud to approximate equilibrium, and therefore, right
now,  the ionization front would probably not be travelling towards the protostar but stalled. This does not
affect the main conclusion of the work; namely, the fact that the morphology of the L1206 outflow is the
result of photoionization from the H{\sc ii} region.}. Following Morgan et
al.~(\cite{morgan04}), we have estimated the internal molecular pressure, $P_{\rm m}=\sigma^2\rho_{\rm m}$, 
where $\sigma$ is the velocity dispersion, which can be given in terms of the observed line width $\Delta v$
as $\sigma^2=(\Delta v)^2/(8\ln 2)$, and $\rho_{\rm m}$ is the density of the molecular gas. The velocity
dispersion has turbulent and thermal contributions. However, as the clouds are composed mostly of cold gas,
one would expect the thermal contribution to be almost negligible. Sugitani et al.~(\cite{sugitani89})
derive $\rho_{\rm m}=2.1\times10^4$~cm$^{-3}$ from $^{13}$CO observations, but they do not give the value of
the observed line width. Thus, using the $^{13}$CO line width derived by Ridge et al.~(\cite{ridge03}), we
obtained $P_{\rm m}/k_{\rm B}\simeq 3.7\times10^6$~cm$^{-3}$ ${\rm K}$. If ones uses $\rho_{\rm m}$  derived
from the observations of Ridge et  al.~(\cite{ridge03}) by assuming spherical symmetry of the cloud, the
values of $P_{\rm m}/k_{\rm B}$ are $1.8\times10^6$ and $5.0\times10^6$~cm$^{-3}$ ${\rm K}$, for $^{13}$CO
and C$^{18}$O, respectively. In summary, the internal molecular pressure ranges  from 1.8 to
$5.0\times10^6$~cm$^{-3}$ ${\rm K}$, depending on the values of the density and line width adopted, but in
any case it is always smaller than the IBL pressure. Thus, the cloud seems to be underpressured with respect
to the IBL, and therefore, one might expect photoionization shocks and the ionization front to be
propagating into the cloud interior, compressing, heating, and photodissociating the molecular gas.
Definitively, the possibility exists that photodissociation is responsible of the morphology of the
molecular outflow, and therefore, that the H{\sc ii} region is actually eating the outflow.

\begin{figure}
\centerline{\includegraphics[angle=-90,width=8cm]{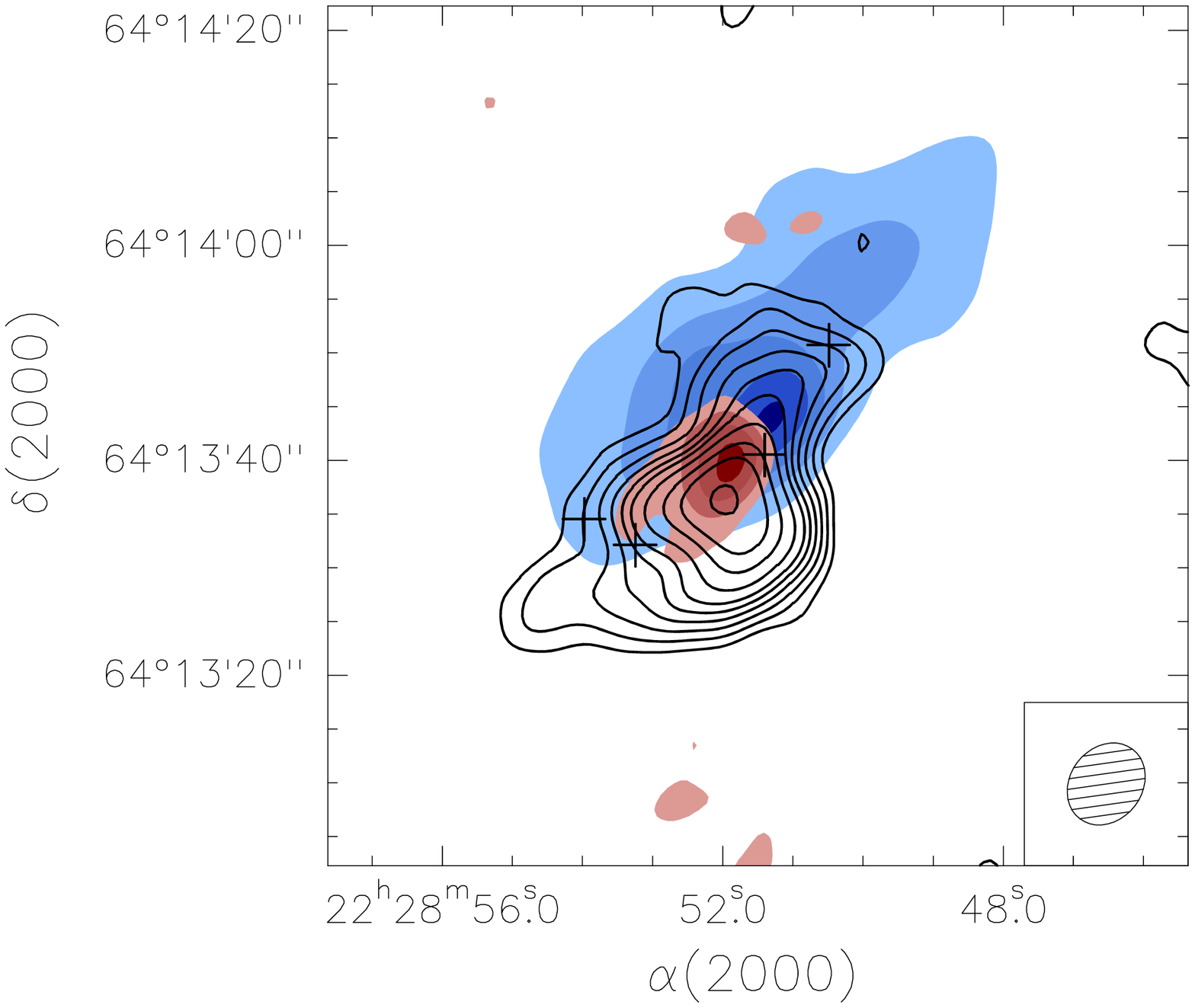}}
\caption{Overlap of the intensity map of the HC$_3$N~(\jdo) emission ({\it contours}) integrated over the
velocity interval ($-10.3$,$-9.7$)~\kms\ on the CO~(\juz) blueshifted emission ({\it blue colours})
integrated over the ($-19.5$,$-13.5$)~\kms\ velocity interval, and the redshifted emission ({\it red
colours}) integrated over the ($-8.5$,$-2.5$)~\kms\ velocity interval. Contour levels are 0.12, 0.2, 0.28,
0.36, 0.44, 0.52, 0.64, 0.80, 0.96, and 1.2\jykms ({\it contours}), 1.5, 5.5, 9.5, 13.5, and 17.5\jykms\
({\it blue colours}), and 0.54, 1.08, 1.62, and 2.16\jykms ({\it red colours}). The crosses show the
positions of the 2.7~mm continuum sources detected in the region.  The synthesized beam is drawn in the
bottom right corner.}
\label{enha}
\end{figure}

\subsection{HC$_3$N enhancement}
\label{interact}

\begin{figure}
\centerline{\includegraphics[angle=-90,width=8cm]{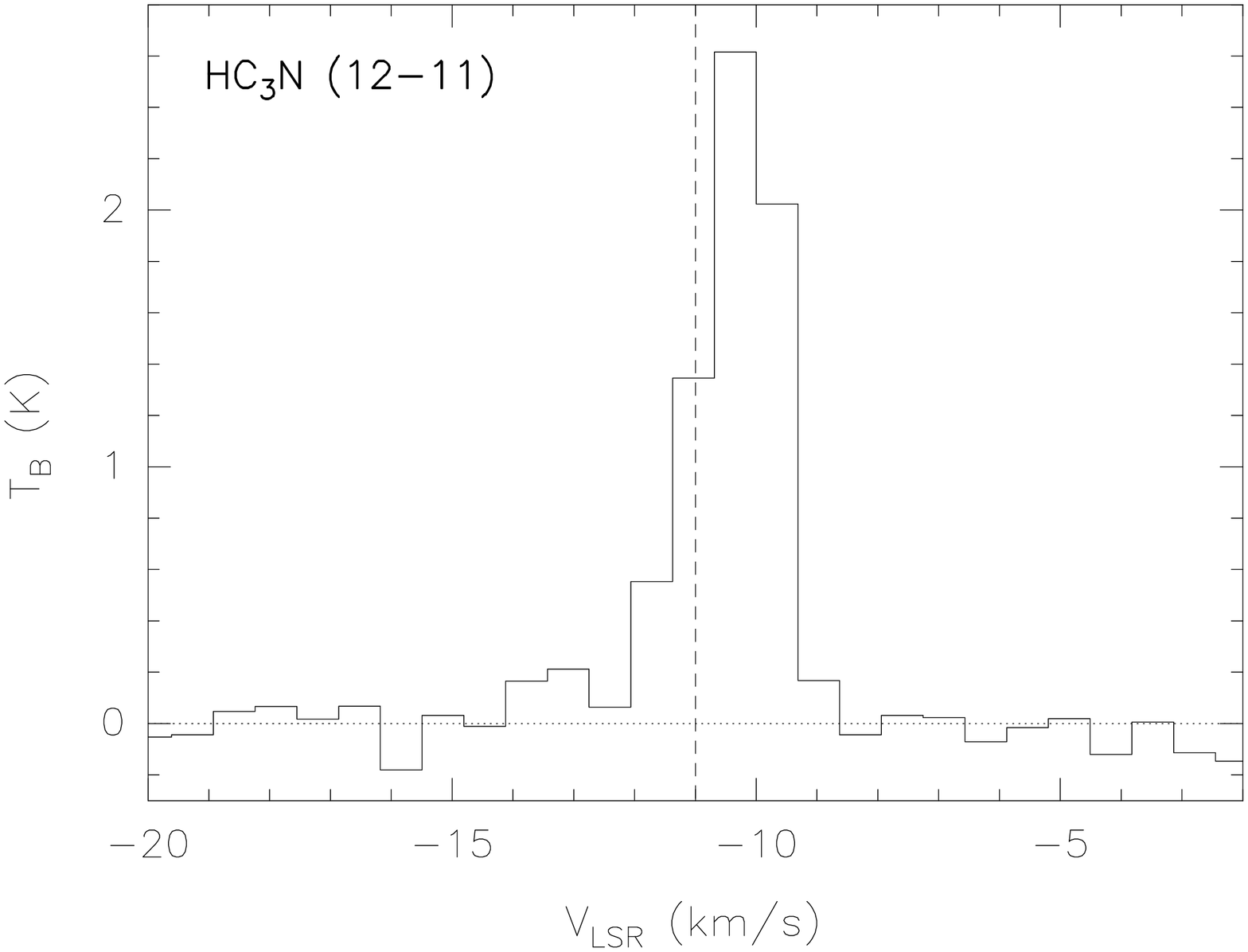}}
\caption{HC$_3$N (\jdo) spectrum obtained at the position of the southern redshifted peak seen at velocities of 
$-9.7$ and $-10.3$~\kms\ (see Sect.~\ref{interact}). The dashed vertical line indicates the systemic velocity of
$-11$~\kms.}
\label{hc3n_redpeak}
\end{figure}

As already mentioned in Sect.~\ref{hc3n}, the HC$_3$N emission at velocities of $-11.0$ and $-11.7$~\kms, which are
close to the cloud velocity, is very compact and traces the same material as the continuum emission at 2.7~mm; that
is, sources OVRO~2,  3, and 4 (see Fig.~\ref{hc3n_channel}). On the other hand, at velocities of $-9.7$ and
$-10.3$~\kms, which are redshifted with respect to the cloud velocity, the HC$_3$N emission is elongated in the same
direction as the CO molecular outflow and peaks south of OVRO~2 (see Figs.~\ref{hc3n_channel} and \ref{enha}). As
already done in Sect.~\ref{gas_emission} toward OVRO~2, we estimated the HC$_3$N fractional abundance toward the red
lobe by integrating the gas and the continuum emission in the same area, and assuming that the mass of the gas derived from 
HC$_3$N is the same as that derived from the continuum. The maximum HC$_3$N fractional abundance 
southwards of OVRO~2 is (2--3)$\,\times10^{-10}$, which compared to $7\times10^{-11}$ estimated toward OVRO~2, 
indicates that the emission is enhanced toward the south. The correlation between the dense gas and the
molecular outflow suggests that the latter could be responsible of the southern HC$_3$N enhancement as a result
of shocks that  would compress and heat the dense gas. Shock-enhancement of the HC$_3$N abundance has also been
detected toward L1157 (Bachiller \& P\'erez-Guti\'errez~\cite{bachiller97}; Beltr\'an et al.~\cite{beltran04a}).
This hypothesis seems to be supported by the HC$_3$N spectrum at the position of the redshifted peak (see
Fig.~\ref{hc3n_redpeak}), which is very different from the spectrum at the position of OVRO~2
(Fig.~\ref{lines}).  Note, however, that the velocity of the HC$_3$N gas is somewhat lower than that of the
CO emission at the same location. This suggests that high velocity material has a lower density, and would thus
not be seen in HC$_3$N.

One should notice that, although the HC$_3$N is well correlated with the molecular outflow, the enhancement
of HC$_3$N is mainly found toward the southern redshifted lobe instead of toward the stronger northern blueshifted
one, as one would expect. Furthermore, the peak is found for redshifted velocities with respect to the
cloud. One possible explanation for this is that, as mentioned in the previous section, the region
southeastern of OVRO~2 is closer to the bright rim and the ionization front, located probably in the
background. Therefore, it is possible that the shock front preceding the ionization front has compressed and
heated the neutral gas, helping to enhance the HC$_3$N abundance. This would support our hypothesis that the
photoionization front that is coming after the shock front is the responsible of the weakness and small size of
the redshifted southern lobe. It is worth noticing that once the photoionization front will reach the
position where the HC$_3$N emission is found, there will be a decrease of its abundance as HC$_3$N
will be dissociated by the UV photons (e.g.\ Mon R2: Rizzo et al.~\cite{rizzo05}).

\section{Conclusions}

We studied with the OVRO Millimeter Array the dust and gas emission at 2.7~mm toward \iras, an intermediate-mass
source embedded in the core of the bright-rimmed cloud L1206. 

The 2.7~mm continuum emission has been resolved into four sources, OVRO~1, OVRO~2, OVRO~3, and OVRO~4. The
strongest source at millimeter wavelengths is OVRO~2, which is most likely the YSO associated with \iras, and it
is probably the driving source of the CO molecular outflow detected in the region.  The millimeter emission of
OVRO~2 shows two components, a centrally peaked source, which has a diameter of $\sim$3200 AU at the 50\% of the
dust emission peak, plus an extended and quite spherical envelope, which has a size of $\sim$11800 AU. The mass
of OVRO~2, which has been estimated from the dust continuum  emission, is 14.2~$M_{\odot}$. The dust emission
morphology and properties of OVRO~2, which are consistent with those of low-mass counterpart protostars,
indicate that it is an extremely young, deeply embedded intermediate-mass protostar, probably in transition
state between Class~0 and I. The sources OVRO~3 and 4 have a similar deconvolved size, $\sim$5000~AU, and mass, 
$\sim$2~$M_{\odot}$, and are probably low-mass protostars.The other source in the region, OVRO~1, could be
dusty material entrained by the outflow detected in the region.

The CO~(\juz) observations have revealed a very weak southeastern redshifted outflow lobe, which is only visible at low
outflow velocities, and a much stronger northwestern blueshifted lobe, which extends to very high velocities. This
collimated molecular outflow is elongated in a direction with PA $\simeq 140\degr$, and it is clearly centered at the position of
OVRO~2. The properties of the outflow are consistent with those of the outflows driven by low-mass YSOs (Bontemps
et al.~\cite{bontemps96}; Anglada~\cite{anglada96}). 

The internal molecular pressure for the bright-rimmed cloud BRC~44 ranges  from 1.8 to $5.0\times10^6$~cm$^{-3}$ ${\rm
K}$, and is smaller than the IBL pressure, $P_{\rm i}/k_{\rm B}$, which
is $8.5\times10^6$~cm$^{-3}$ K (Morgan et al.~\cite{morgan04})\footnote{See footnote \#2.}. This suggests that the cloud is underpressured
with respect to the IBL, and therefore, one might expect photoionization shocks and the ionization front to be
propagating into the cloud interior, compressing, heating, and photodissociating the molecular gas. Therefore,
photodissociation could explain the weakness of the redshifted lobe and the morphology of the molecular outflow,
drawing an scenario where the H{\sc ii} region would be, in fact, eating the redshifted outflow lobe.

The spatial correlation between the outflow and the elongated dense protostellar material, as traced by
HC$_3$N~(\jdo), suggests an interaction between the molecular outflow and the protostellar envelope. Shocks
produced by the molecular outflow, and possibly by the shock front preceding the photoionization front coming from
the bright rim, would be responsible of the southern enhancement of the HC$_3$N abundance for redshifted outflow
velocities.

The C$^{18}$O abundance relative to molecular hydrogen estimated
toward OVRO~2 is $3\times10^{-8}$, a value $\sim$6 to 13 times lower that typical fractional abundances
estimated toward molecular clouds. The most plausible explanation for such a difference is CO
depletion toward OVRO~2. The HC$_3$N abundance relative to molecular hydrogen estimated
toward OVRO~2 is $7\times 10^{-11}$.

\begin{acknowledgements}

It is a pleasure to thank the OVRO staff for their support during the observations. MTB, JMG, and RE are
supported by MEC grant AYA2005-08523-C03.

\end{acknowledgements}

\end{document}